\documentclass[a4paper,fleqn]{cas-dc}

\usepackage[square, numbers]{natbib}
\usepackage{gensymb}
\usepackage{graphicx, subcaption}
\usepackage{lipsum}
\usepackage{float}
\usepackage{multirow}
\usepackage{array}
\usepackage{soul}

\def\tsc#1{\csdef{#1}{\textsc{\lowercase{#1}}\xspace}}
\tsc{WGM}
\tsc{QE}
\tsc{EP}
\tsc{PMS}
\tsc{BEC}
\tsc{DE}

\begin{document}
\let\WriteBookmarks\relax
\def\floatpagepagefraction{1}
\def\textpagefraction{.001}

\shorttitle{Fire Safety Journal}

\shortauthors{Vijayananda V. Devananda \& Tarek Echekki}

\title [mode = title]{Numerical investigation of gravity effects on backdraft phenomena in an enclosure with varying opening geometries}                      



%
\author[]{Vijayananda V. Devananda}[
                        ]





\credit{Conceptualization of this study, Investigation, Methodology, Software, Writing-original draft, Writing-review \& editing}

\affiliation[]{organization={Department of Mechanical and Aerospace Engineering, North Carolina State University},
    city={Raleigh},
    postcode={27695}, 
    state={NC},
    country={USA}}

\author[]{Tarek Echekki}[orcid=0000-0002-0146-7994]

\cormark[1]

\ead{techekk@ncsu.edu}

\credit{Conceptualization of this study, Methodology, Supervision, Writing-original draft, Writing-review \& editing}


\cortext[cor1]{Corresponding author}



\begin{abstract}
In this study, we investigated numerically the backdraft phenomenon under different gravity conditions and 4 openings using the Fire Dynamics Simulator (FDS) code. Four different opening geometries are studied under ten different gravity conditions. As demonstrated by an earlier study in our group, backdraft is established under different gravity conditions from 0.1 $g$ to 1 $g$. The rate at which oxygen reaches an ignition block in the enclosure in the presence of gravity currents plays an important role in the onset of ignition, the subsequent backdraft formation, and the maximum pressure built inside the enclosure before the onset of backdraft. This role also explains why these effects are different under different openings. We observe that the gravity strength affects the ignition time and the onset of backdraft nonlinearly. Moreover, the smoke exiting the enclosure is a precursor for the onset of backdraft, albeit with a short warning, allowing people on the outside to undertake necessary precautions. The effect of backdraft in the form of heat flow and impact force at the exit is also studied. It is found that the effect of heat flow is more severe than that of the impact force.
\end{abstract}

\begin{highlights}
\item Simulations of backdraft under different gravity conditions and opening geometries are implemented.
\item The Fire Dynamics Simulator (FDS) is used in simulations with methane as fuel and with finite-rate chemistry.  
\item Backdraft is present for all opening geometries and gravity conditions.
\item The different opening geometries exhibit different ignition delay times, maximum pressures inside the enclosures, and different times for the onset of backdraft.
\item Smoke release before backdraft may provide up to a few seconds of precursor for backdraft.
\item Heat from the expanding combustion gases poses more hazard than the impact force of these gases.
\end{highlights}


\maketitle

\section{Introduction}\label{introduction}
Backdraft is the deflagrative burning of a combustible mixture of gases in an oxygen-depleted enclosed space~\cite{fleischmann1993backdraft}, \cite{fleischmann_exploratory_1993},  \cite{Fleischmann_numerical}. Oxygen may rush through an opening of an enclosure due to a gravity current, facilitating conditions for the reignition of the rich fuel mixture. The deflagration exits the opening with rapid heat release. Though rare, backdraft poses a severe risk. 

Gravity plays an important role in driving the flow inside the enclosure ~\cite{fleischmann1993backdraft}, \cite{fleischmann_exploratory_1993}, \cite{weng2005} and its effect on the onset of gravity currents. As humans explore interplanetary travel and the construction of settlements outside our planet, it is important to understand this role under different gravity conditions.

Ashok and Echekki~\cite{ashok_numerical_2021} investigated numerically the effects of gravity on the onset and growth of backdraft phenomena under different fractions of Earth gravity. They showed that even at reduced gravity conditions, backdraft phenomena are present and can exhibit similar growth characteristics as backdraft phenomena at Earth gravity.

The study by Ashok and Echekki~\cite{ashok_numerical_2021} was based on one opening geometry, which corresponds to a vertical opening along the height of the enclosure. This configuration was chosen to emulate the experiment of Weng and Fan~\cite{weng_critical_2003}. In both studies, the working fuel was methane, and its chemistry was modeled using a single-step reaction. The rate parameters for this reaction in the study of Ashok and Echekki~\cite{ashok_numerical_2021} were tuned to predict the onset of backdraft in the experiment.

The study by Ashok and Echekki~\cite{ashok_numerical_2021} motivates the present study and attempts to address 3 questions spurred by the earlier work. The first question is related to the role of the opening placement and size given the importance of the gravity current in drawing oxygen to the compartment.

Several computational studies have been carried out to evaluate the effect of the opening geometry and placement on backdraft phenomena. Weng et al.~\cite{weng2005} investigated 8 different opening geometries with different opening areas. They found that for the same area of the opening, the gravity current velocity is the lowest for the opening closer to the enclosure ceiling and highest for the opening at the bottom. They also observed that the gravity current velocity decreases with decreasing opening area. 

In their work, Ferraris et al.~\cite{Ferraris_2009} found that the ignition time is directly affected by the vent opening area. They observed that small opening areas located far away from the floor lead to larger ignition delays. They also report a significantly higher maximum pressure for an opening with a smaller area than the larger one also implying a larger ignition delay will lead to higher maximum pressures.

Myilsamy et al.~\cite{myilsamy_large_2019} observed that the peak pressure in the enclosure is directly related to the amount of air supplied from the opening and the mixing status of fuel and air. They observed a higher peak pressure for opening with a higher ignition delay and consequently a larger air entrainment time.

The second question is related to the potential presence of precursors, such as smoke release, that can help provide sufficient warning to anticipate the onset of a backdraft. The third question is related to the impact of backdraft on occupants or firefighters who stand close to the exit of the deflagration wave. Is the wave impact primarily thermal resulting in burns or is the gas expansion powerful enough to create a sudden impact force?

This work is implemented to address the above questions. As in the Ashok and Echekki~\cite{ashok_numerical_2021} work it is primarily based on the experimental setup of Weng and Fan~\cite{weng_critical_2003}. With the same enclosure configuration, four different opening geometries are studied. The first opening geometry is identical to the setup of Ashok and Echekki~\cite{ashok_numerical_2021} and is used as the reference case.

The paper is organized as follows. In Sec.~\ref{methods} the numerical setup and post-processing procedures are presented. Results are presented and discussed in Sec.~\ref{results}. These results are summarized in the last conclusions Sec.~\ref{conclusions}.

\begin{figure*}[h!]  
	\centering
	\includegraphics[scale=.5]{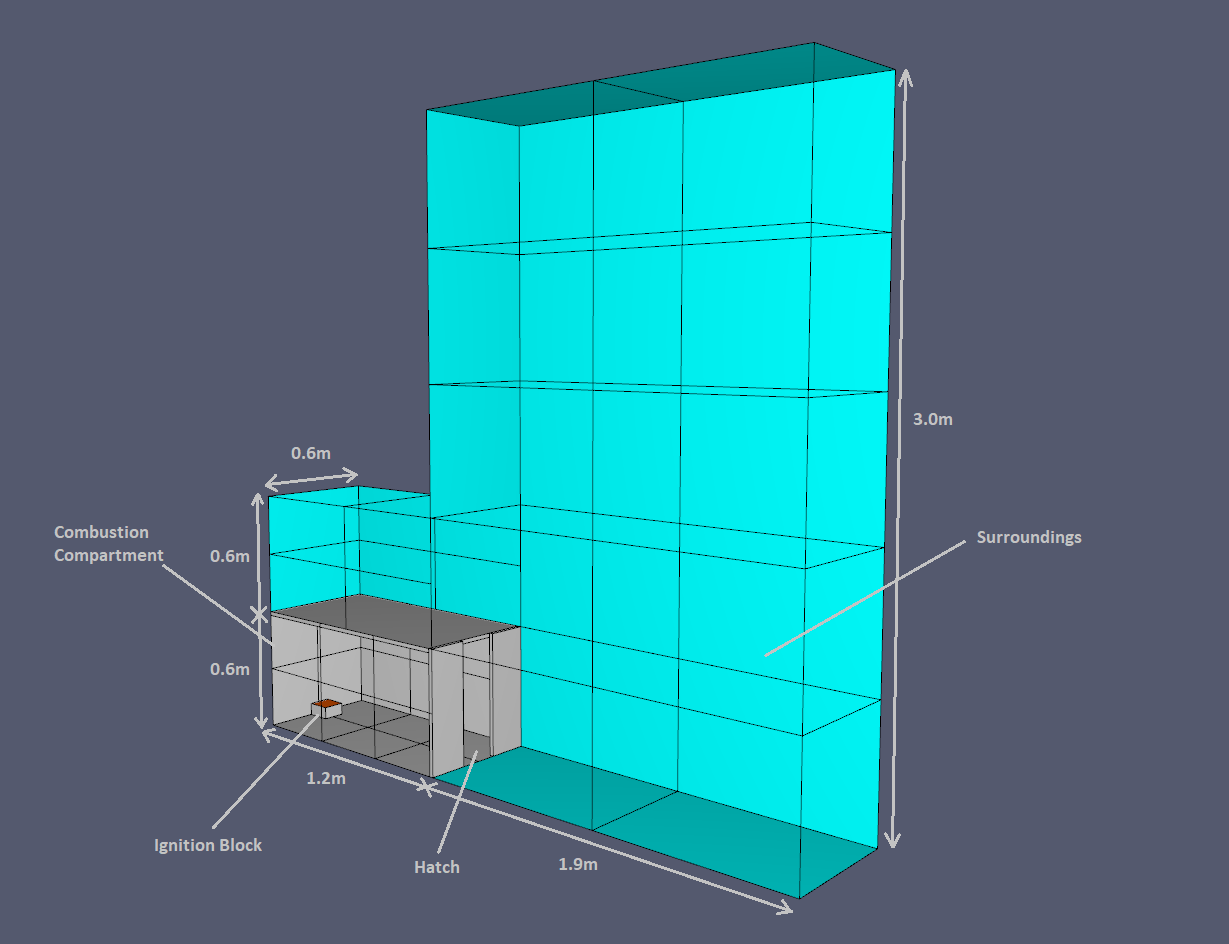}
	\caption{Computational domain and geometry of model 1 with a vertical opening.}
	\label{comp_geometry}
\end{figure*}

\section{Numerical method and setup}\label{methods}
The NIST Fire Dynamics Simulator (FDS) version 6.8~\cite{mcgrattan_fire_2013} was used for simulating the backdraft with finite-rate chemistry. FDS is a low-Mach number CFD code for modeling fire-driven flows. In the present study, we rely on its large-eddy simulation (LES) formulation to capture the dynamics of the flow and the fire growth. The graphical user interface PyroSim~\cite{pyrosim_userm}, \cite{pyrosim_results_userm} was used for problem setup, including the generation of the FDS input file, data visualization, and post-processing of the results. Post-processing was also done using a visualization program called Smokeview \cite{smv_forney}.

\subsection{Problem configurations}
In the present study, we model backdraft using the same enclosure geometry as in Ref.~\cite{ashok_numerical_2021} with four different openings, identified as models 1, 2, 3, and 4. 
Figure~\ref{comp_geometry} shows the overall model geometry, which includes the enclosure (shown in gray) and the surrounding domain. The opening configuration corresponds to model 1, which is identical to the configuration used by Ashok and Echekki~\cite{ashok_numerical_2021} and Weng and Fan~\cite{weng_critical_2003}.

\begin{figure*} 
	\centering
	\includegraphics[scale=.39]{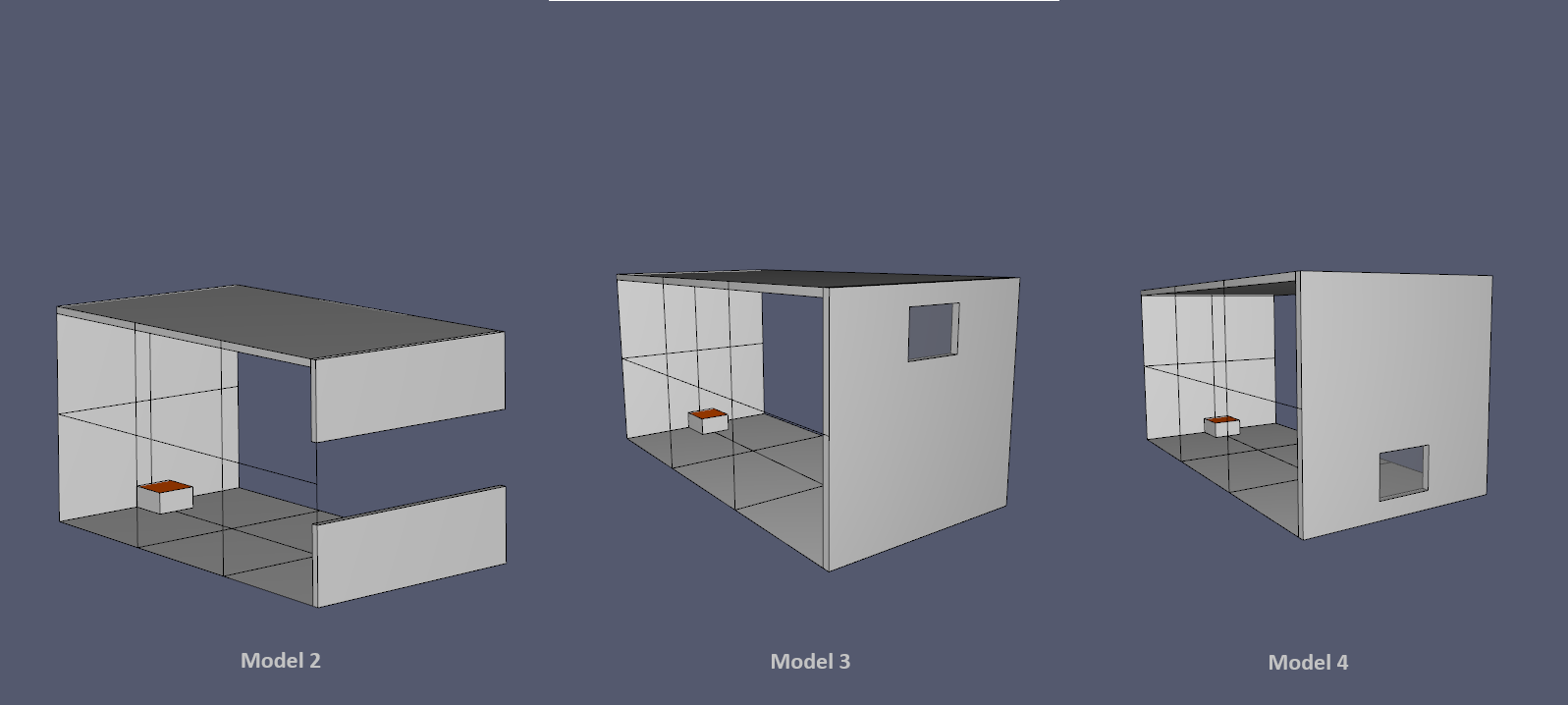}
	\caption{Opening geometries of models 2, 3, and 4.}
	\label{opening_geometry}
\end{figure*}

Figure~\ref{opening_geometry} shows the portion of the computational domains that include only the enclosure with the different opening geometries for models 2, 3 and 4. The enclosure has dimensions of  1.2 m $\times$ 0.6 m $\times$ 0.6 m. The thickness of the wall is 0.02 m. The surrounding volume directly above the compartment has dimensions of 1.9 m $\times$ 0.6 m $\times$ 3 m. There is an ignition block of dimensions 0.12 m $\times$ 0.12 m $\times$ 0.06 m located along the wall and on the ground opposite to that of the opening. This location is the same as that in the experiment where the ignition source is located. 

The opening for model 1 extends from the bottom to the top of the enclosure floor and ceiling. The dimensions of this opening are 0.2 m $\times$ 0.6 m. Model 2 has a horizontal opening extending from the left to the right walls with dimensions 0.6 m $\times$ 0.2 m. The openings for models 3 and 4 are smaller with dimensions 0.18 m $\times$ 0.12 m, in the horizontal and vertical directions. The opening in model 3 is placed toward the top of the enclosure, and the opening in model 4 is placed toward the bottom of the enclosure.

It is important to note why we have adopted these 4 openings. Models 1 and 2 are characterized by larger areas, which, in principle, should facilitate the inflow of oxygen and the outflow of burned gases. The choice of models 3 and 4 is motivated by the need to understand the placement of the openings and its impact on the inflows into and the outflows out of the compartment. We anticipate that the flow path and the area of the opening will play a role in anticipating or delaying the ignition process and the subsequent backdraft formation.

\subsection{Simulation implementation}
One of the main challenges of the current work was the numerical modeling of the problem. A direct numerical simulation (DNS) is not feasible owing to unrealistic computational costs. Also, simulating the experimental conditions from the beginning that involved fire extinction, re-ignition and propagation adds unnecessarily significant computational overhead. As mentioned in the work of Weng and Fan~\cite{weng_critical_2003}, the hatch of the enclosure remains closed for a specific amount of time during which the fire extinguishes due to lack of oxygen. The hatch is then opened. Weng and Fan recorded the conditions inside the compartment when the hatch was opened. This is taken as the starting point of the simulation in this work. We believe that this is sufficient given that the backdraft is initiated only after the hatch is opened. This starting point also avoids simulating the costly extinction and re-ignition process.

The entire domain is discretized with a resolution of 2 cm. This results in a total of 535,500 cells that are cuboid in shape. A grid convergence study was conducted using a resolution of 1 and 2 cm. The maximum pressure and heat release rates were compared between the two cases. The maximum gauge pressures for the 2 cm and 1 cm grid models are 8 Pa and 11.64 Pa respectively. This is not a significant variation considering the value of the total thermodynamic pressure. The maximum heat release rate in the enclosure is 425.54 kW and 414.94 kW respectively for the 2 cm and 1 cm grid. It was thus concluded that the model with a 2 cm grid was adequate for this study rather than a finer and more computationally expensive model. Taking advantage of FDS' parallel processing capabilities, the problem was divided into 26 subdomains managed each by one CPU. This enabled faster generation of the simulation databases that span different conditions and realizations.

FDS by default uses a mixing-controlled combustion model, which is based on an infinite-rate chemistry model. However, the backdraft phenomenon cannot be accurately captured using the mixing-controlled combustion model as shown by Park et al. \cite{park_computational_2017}, \cite{park_J.-W._study_2013} in their work. Thus, a finite-rate combustion model must be adapted to simulate the backdraft experiments. The following single-step chemical reaction is used in this work:
\begin{equation} \label{reaction}
   \mathrm{CH_4 + 1.9 \ O_2 \rightarrow 0.9 \ CO_2 + 2 \ H_{2}O + 0.1  \ C_s}
\end{equation}
The differentiating factor from the work of Ashok and Echekki~\cite{ashok_numerical_2021} is that soot in the form of carbon was added to the chemical reaction. This addition is made to investigate whether soot release through the enclosure opening can serve as a precursor for the onset of backdraft. The coefficient of soot 0.1 in the chemical reaction is chosen arbitrarily. The fuel reaction rate is expressed as follows:
\begin{equation} \label{rate_of_reac}
    \mathrm{\frac{d[CH_4]}{dt}}= k\  \mathrm{[CH_4] [O_2]^{1/2} \  \  \ mol/cm^3/s}
\end{equation}
where the symbol "[ ]" indicates the concentration of the species in $\mathrm{mol/cm^3}$. The rate constant for the single-step chemical reaction (\ref{reaction}) is given by Arrhenius law as shown below:
\begin{equation} \label{rate_constant}
k = A \ \exp\left( - \frac{E_a}{R T} \right)\ \ \  \ \mathrm{\left(cm^3/mol\right)^{1/2}\cdot s}
\end{equation}
where $k$ is the rate constant for the reaction (\ref{reaction}), $A$ is the pre-exponential factor in $\mathrm{(cm^3/mol)^{1/2}\cdot s}$, $E_a$ is the activation energy in $\mathrm{J/mol}$, $R$ is the universal gas constant (8.314 $\mathrm{J/K-mol}$), and $T$ is the absolute temperature in Kelvins. The frequency factor $A$ is taken as $\mathrm{1.49 \times 10^{12} (cm^3/mol)^{1/2}.s}$ and $E_a$ to be 107,460 $\mathrm{J/mol}$. As done by Ashok and Echekki~\cite{ashok_numerical_2021}, these values have been modified from what is reported in Westbrook and Dryer~\cite{westbrook_simplified_1981} to accommodate for the fact that LES spatially averages the temperature and species concentration.

Since there is an additional component in the chemical reaction i.e. soot, the activation energy was optimized to yield a pressure value close to that obtained in the experiment in model 1 under normal gravity conditions. The critical flame temperature is set to the default value of 1507\degree C for methane-oxygen reaction. The auto-ignition temperature for methane is set at 150\degree C, a value much lower than the actual one, to avoid spurious ignitions away from the heat source. 

In the experiment conducted by Weng and Fan~\cite{weng_critical_2003}, it is noted that there is a formation of two separate regions within the enclosure with different temperatures just after the hatch is opened. Mass fractions of various gases are also recorded at this point. As stated earlier, the simulation in this work starts from this point. Weng and Fan~\cite{weng_critical_2003} conducted eight different backdraft experiments. Each experiment has varying ambient temperature, fuel flow rate, and time of flow. Out of the eight experiments, the eighth experiment is recreated in this work. The initial conditions inside the enclosure are shown in Table~\ref{Initial conditions}.
\begin{table}[H]
    \centering
\caption{Initial conditions in the enclosure.}
\label{Initial conditions}
    \begin{tabular}{|c|c|c|} \hline 
         Quantities&  Upper layer& Lower layer\\ \hline \hline
         Height (m)&  0.32& 0.28\\ \hline 
         Temperature (\degree C)&  103& 67\\ \hline 
         $\mathrm{N_2}$ mass fraction (kg/kg)&  \multicolumn{2}{|c|}{0.6994}\\ \hline
         $\mathrm{O_2}$ mass fraction (kg/kg)&  \multicolumn{2}{|c|}{0.146}\\ \hline 
         $\mathrm{CH_4}$ mass fraction (kg/kg)&  \multicolumn{2}{|c|}{0.1224}\\ \hline 
         $\mathrm{CO_2}$ mass fraction (kg/kg)&  \multicolumn{2}{|c|}{0.021}\\ \hline 
         CO mass fraction (kg/kg)&  \multicolumn{2}{|c|}{0.0012}\\ \hline 
         Soot mass fraction (kg/kg)&  \multicolumn{2}{|c|}{0.01}\\ \hline
    \end{tabular}    
\end{table}
An arbitrary soot mass fraction of 0.01 kg/kg is added to the enclosure as seen in the initial conditions. This accounts for the fact that during the time the fire goes from ignition to extinction, the fuel burnt in a rich stoichiometric ratio that could have led to incomplete combustion and generation of smoke. The background pressure in the domain is set as $1.01325 \times 10^5$ Pa. All reported values of pressures in this work are gauge pressures. The ambient temperature outside the enclosure is set at 20\degree C where the ambient air composition (in mass fraction) of nitrogen, oxygen, water vapor, and carbon dioxide is 0.76274 kg/kg, 0.23054 kg/kg, 0.00626 kg/kg, and 0.00046 kg/kg, respectively. 

The walls of the enclosure are modeled as adiabatic, which is a close approximation of the experiment. The rest of the computational domain was modeled as open to the surroundings. The top face of the ignition block is modeled to have a constant temperature of 1500\degree C. 

Random perturbations are introduced into the simulation to construct ensembles of backdraft events under similar conditions. In FDS, a parameter called NOISE\textunderscore VELOCITY introduces a small amount of ``noise'' or initial flow velocity into the flow field. In principle, this noise can also be used to generate different realizations that correspond to the same statistics. By default, this value is 0.005 m/s. Randomization in the simulation can be achieved by varying the value of the NOISE\textunderscore VELOCITY. In this work, a set of three simulations were run each with different NOISE\textunderscore VELOCITY values. 

Although the number of such realizations is small, keeping 3 realizations does lessen fluctuations of global quantities presented in this study, including maximum pressure values, ignition and deflagration delay times, the delay times between smoke release and backdraft, and the impact forces and heat flow. The details of the simulations and their corresponding NOISE\textunderscore VELOCITY values are shown in Table~\ref{simulation sets}. 
\begin{table}[H]
    \centering
\caption{Simulation sets and corresponding NOISE\textunderscore VELOCITY values.}
\label{simulation sets}
    \begin{tabular}{|c|c|} \hline 
         Simulation set& NOISE\_VELOCITY (m/s)\\ \hline \hline
         1& 0.005\\ \hline 
         2& 0.004\\ \hline 
         3& 0.006\\ \hline
    \end{tabular}
\end{table}
The total number of simulations in FDS is 120. They correspond to the 4 openings considered and consideration of 10 gravity values from 0.1 $g$ to 1.0 $g$ at an increment of 0.1 $g$ along with 3 realizations of the same conditions.
\subsection{Post-processing of the results}
In addition to the 3D results for velocity and scalar contours at different time increments generated through PyroSim and Smokeview, we also evaluated integral quantities that measure the impact of the backdraft events in the 4 different opening configurations. They are designed to evaluate: 1) the onset of ignition triggered by the increase in O$_2$ mass fraction at the end of the enclosure, 2) the average impact force at the exit plane of the enclosure, and 3) the heat impact at the same exit.  
\begin{figure*}
	\centering
	\includegraphics[scale=.57]{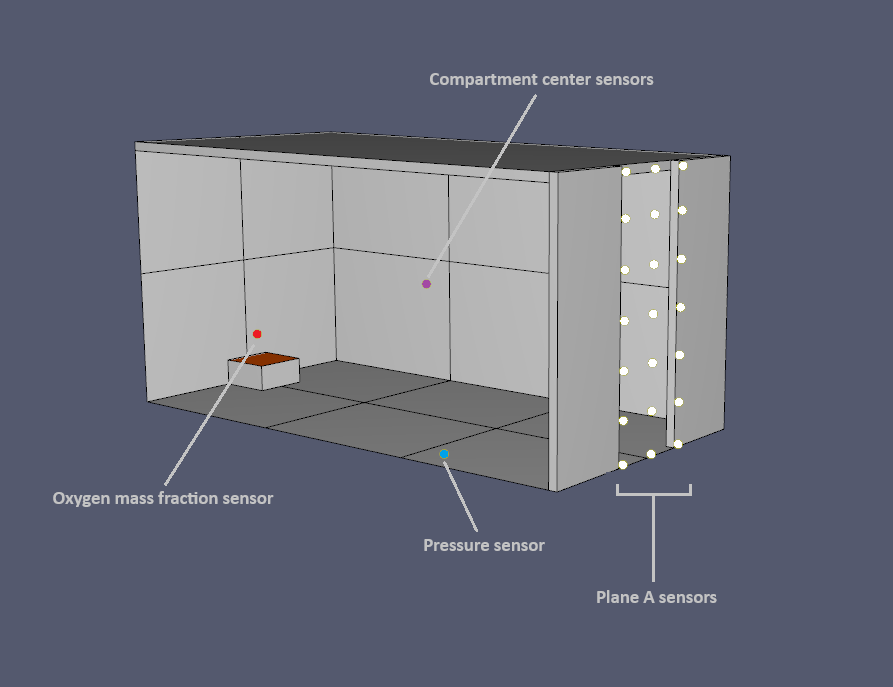}
	\caption{Sensor locations in model 1.}
	\label{sensor_loc}
\end{figure*}
An oxygen mass fraction sensor is located above the ignition block at location (0.04 m, 0.3 m, 0.12 m). It detects the arrival of the gravity current to the farther end of the enclosure. This sensor activates when the oxygen mass fraction rises above 0.2 kg/kg which then activates the ignition block. A pressure sensor is located at (0.9 m, 0.02 m, 0.02 m) that measures the pressure at that point. This sensor is at the same location as in the experiment and many other numerical studies~\cite{ashok_numerical_2021}, \cite{myilsamy_large_2019}, \cite{park_computational_2017}, \cite{park_J.-W._study_2013}. Sensors measuring the mass fraction of various gases taking part in the chemical reaction and soot are placed at the center of the compartment as well as at the centroid of the opening. 

In FDS, there is no direct way to measure the average amount of heat or impact force transferred through an area in a plane. Thus, several sensors measuring flow properties such as density, cell center $u$-velocity (in the horizontal direction), the absolute value of the cell center $u$-velocity, and heat flux are placed on the exit plane ($x = $ 1.2 m) in an area projected by the opening.  In models 1 and 2, there are 21 points in space where these sensors are placed. In the case of models 3 and 4, there are 12 points. Figure~\ref{sensor_loc} shows the location of various sensors in model 1. A tolerance of 0.01 m is provided in those areas where the sensors intersect the compartment and boundaries of the domain to avoid errors. 

The average impact force through the area projected by the opening on the exit plane is calculated using the equation:
\begin{equation} \label{impact_eq}
  \mathrm{\Vec F} = \frac{\sum_{1}^{n} \rho \Vec{u} |\Vec{u}|}{n}\ A \ \ \ \mathrm{N}
\end{equation}
where $\mathrm{\Vec{F}}$ is the average impact force, $\mathrm{\rho}$ is the density of the gases passing through the sensor, $\mathrm{\Vec{u}}$ is the cell-center $u$-velocity of the gases passing through the sensor, and A is the area projected by the opening on the exit plane. $A$ is 0.116 $\mathrm{m^2}$ for models 1 and 2. It is 0.0216 $\mathrm{m^2}$ for models 3 and 4. $\mathrm{n}$ is the number of sensors in a plane. $\mathrm{n}$ is 21 in models 1 and 2. It is 12 in models 3 and 4. The average heat transfer through the area projected by the opening on the exit plane is calculated using the equation:
\begin{equation} \label{heat_flow}
   \mathrm{\Vec{Q}} = \frac{\sum_{1}^{n} \Vec{q''}}{n} \ A \ 
   \ \ \ \ \mathrm{kW}
\end{equation}
where $\mathrm{\Vec{Q}}$ is the average heat flow, and $\Vec{q''}$ is the heat flux passing though the sensor~\cite{mcgrattan_fire_2013}. The sensor measuring the heat flux is called ``gauge heat flux gas'' in FDS and it measures the heat flux using the equation:
\begin{equation} \label{heat_flux}
   \mathrm{\dot{q}''_{gauge}} = \epsilon_{gauge}(\dot{q}''_{inc} - \sigma T^4_{gauge}) + h_c(T_g - T_{gauge}) \ 
   \ \ \ \ \mathrm{kW/m^2}
\end{equation}
where $\mathrm{\dot{q}''_{gauge}}$ is the heat flux measured by the sensor, $\epsilon_{gauge}$ is the emissivity of the water-cooled heat flux gauge, $\dot{q}''_{inc}$ is the incident heat flux in $\mathrm{kW/m^2}$, $\sigma$ is the Stephan-Boltzmann constant $\mathrm{(5.6703 \times 10^{-8} W/m^2K^4)}$, $\mathrm{T_{gauge}}$ is the temperature of the gauge (taken as the ambient temperature), $\mathrm{h_c}$ is the heat transfer coefficient calculated from local gas temperature and velocity, and $\mathrm{T_g}$ is the gas temperature at a virtual flat plate normal to the orientation vector of the sensor (this orientation vector is pointing in the positive x axis). More details and assumptions about the equation \ref{heat_flux} can be seen in reference \cite{mcgrattan_fire_2013}.

\section{Results and discussions}\label{results}
In this section, we present the results of the simulations carried out on the 4 closure openings and the 10 values of the gravity value considered. Global trends, including the maximum pressure, the ignition time, the backdraft formation time, the impact force, and the thermal exposure are averaged over the 3 realizations for each case. Meanwhile, individual contours (e.g. soot and fire, velocity vectors) are shown for one of the realizations, which corresponds to a NOISE\textunderscore VELOCITY equal to 0.006 m/s.

\begin{figure*}[!ht]
 \centering
 \begin{subfigure}[b]{0.48\textwidth}
     \centering
     \includegraphics[scale=0.26, trim={1cm 1cm 1cm 1cm},clip]{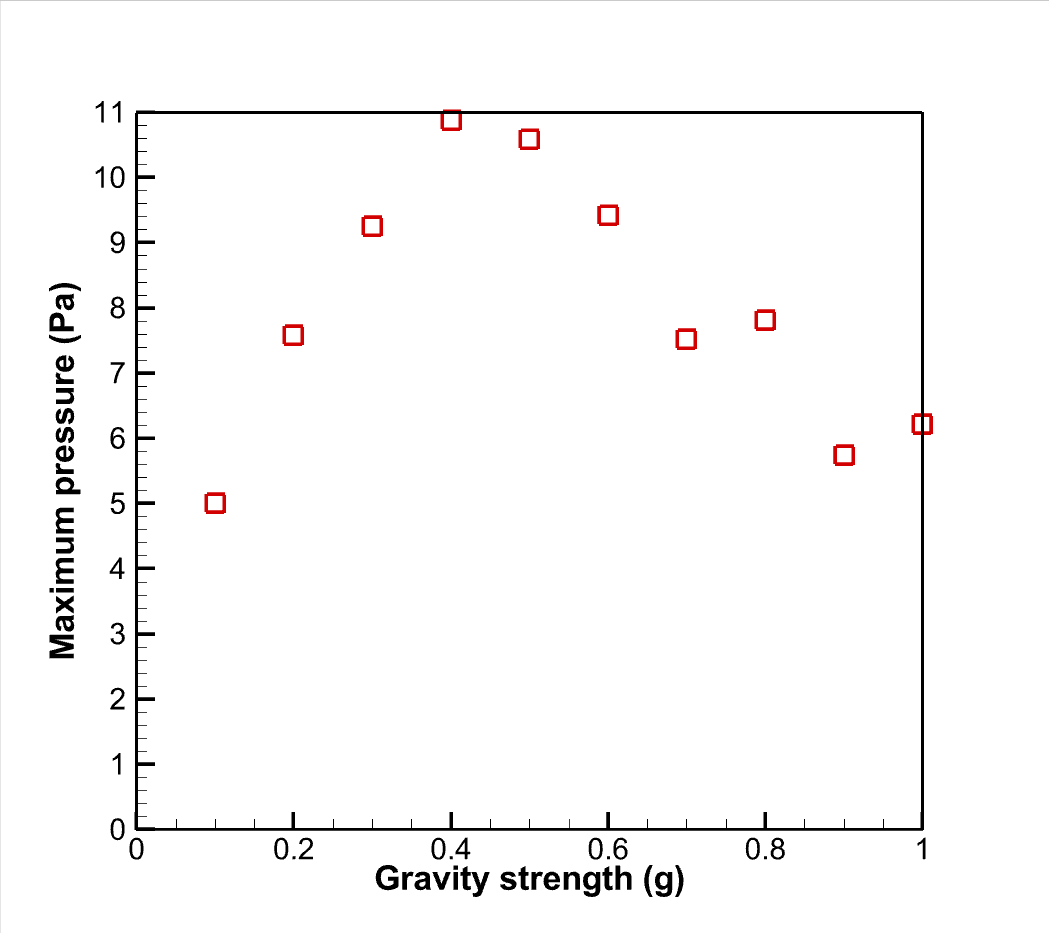}
     \caption{Model 1}
     \label{maxp_model1}
 \end{subfigure}
 \hfill
 \begin{subfigure}[b]{0.48\textwidth}
     \centering
     \includegraphics[scale=0.26, trim={1cm 1cm 1cm 1cm},clip]{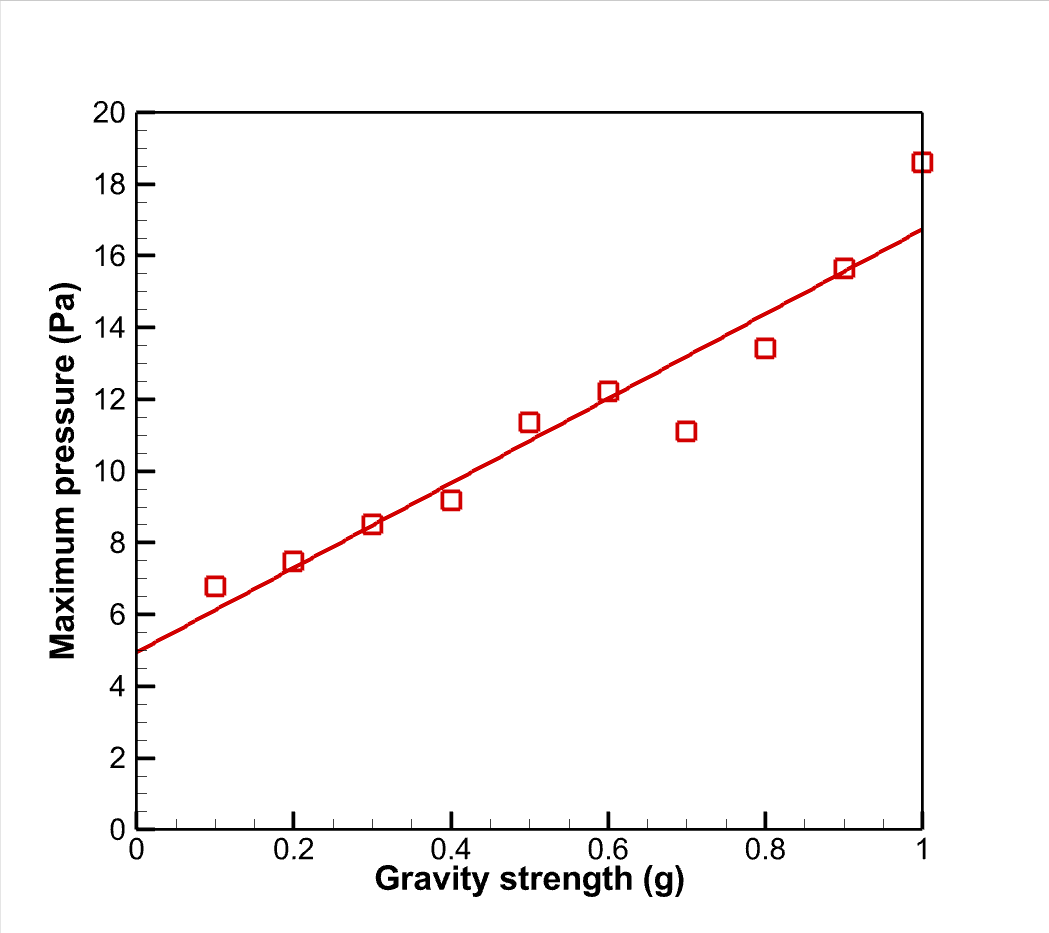}
     \caption{Model 2}
     \label{maxp_model2}
 \end{subfigure}
 \vskip\baselineskip
 \begin{subfigure}[b]{0.48\textwidth}
     \centering
     \includegraphics[scale=0.26, trim={1cm 1cm 1cm 1cm},clip]{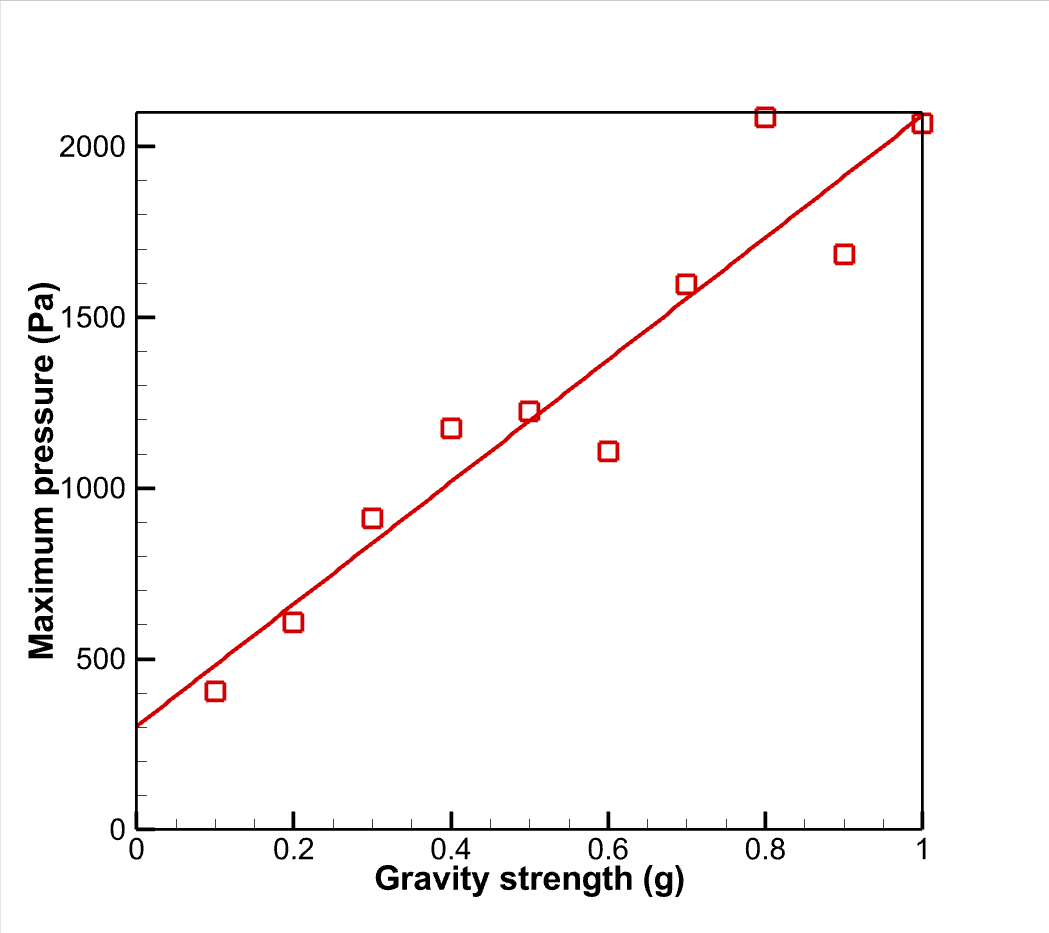}
     \caption{Model 3}
     \label{maxp_model3}
 \end{subfigure}
\hfill
 \begin{subfigure}{0.48\textwidth}
     \centering
     \includegraphics[scale=0.26, trim={1cm 1cm 1cm 1cm},clip]{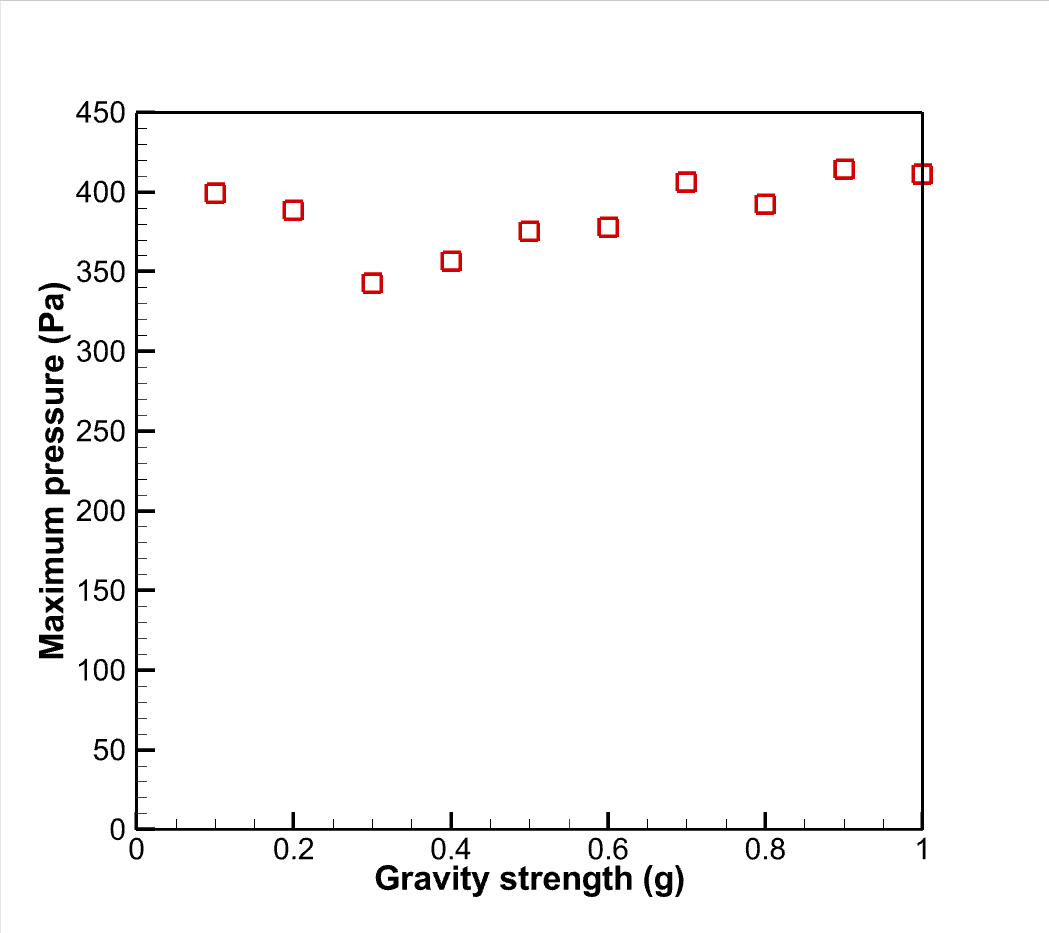}
     \caption{Model 4}
     \label{maxp_model4}
 \end{subfigure}
 
 \caption{Maximum pressure vs. gravity strength for the four models. Red squares are data points, and the line is a linear fit.}
 \label{maxP}
\end{figure*}

Figure~\ref{maxP} shows the variation of the maximum gauge pressure recorded by the pressure sensor with gravity strength. It can be observed that the maximum pressure tends to decrease with gravity strength for all the models except for the first and fourth ones. In model 1, there is no clear trend due to large fluctuations in the flow field. Another observation from these plots is that the maximum pressures are around the same ranges for models 1 and 2. The maximum pressure value in model 3 rises to approximately 2000 Pa under normal gravity conditions. The maximum pressure trend in model 4 is not clear though the maximum pressure values vary in the vicinity of 400 Pa. 

\begin{figure*}[hbt!]  
	\centering
	\includegraphics[scale=0.55]{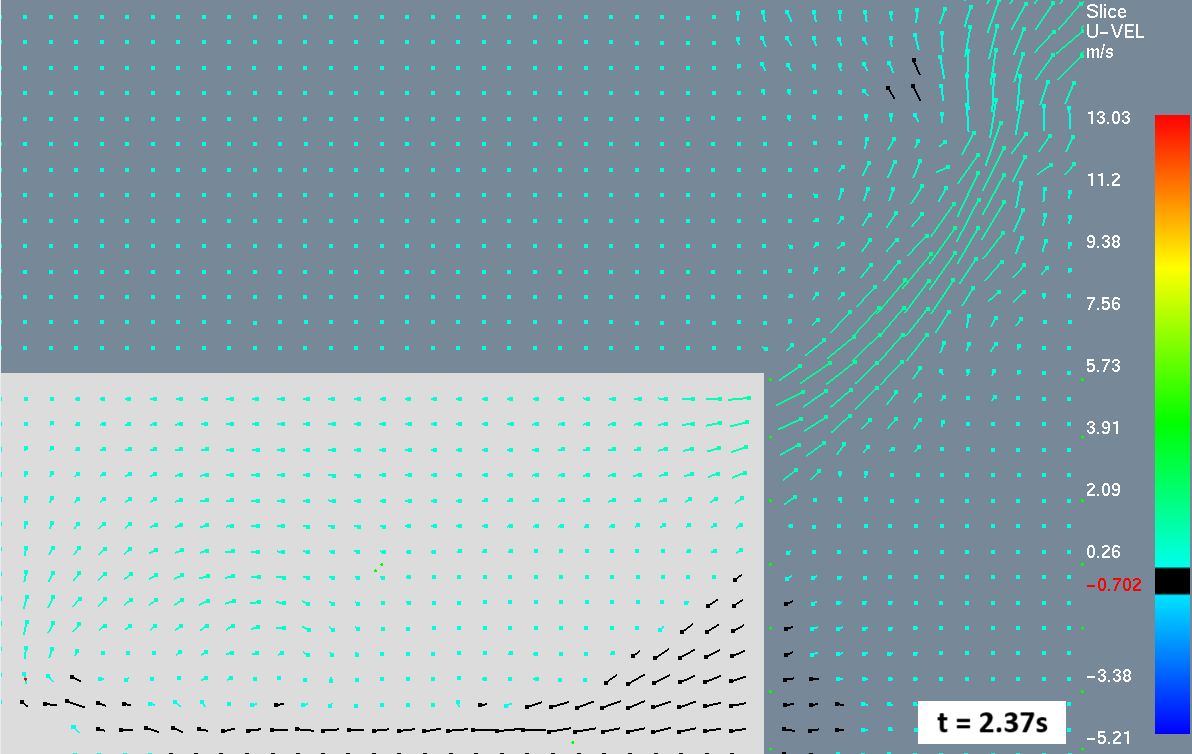}
	\caption{2-D velocity vector contour of model 1 under normal gravity conditions at the time of ignition on the central $y$-plane.}
	\label{inlet_vel_m1}

\vspace{0.5cm}

	\centering
	\includegraphics[scale=0.55]{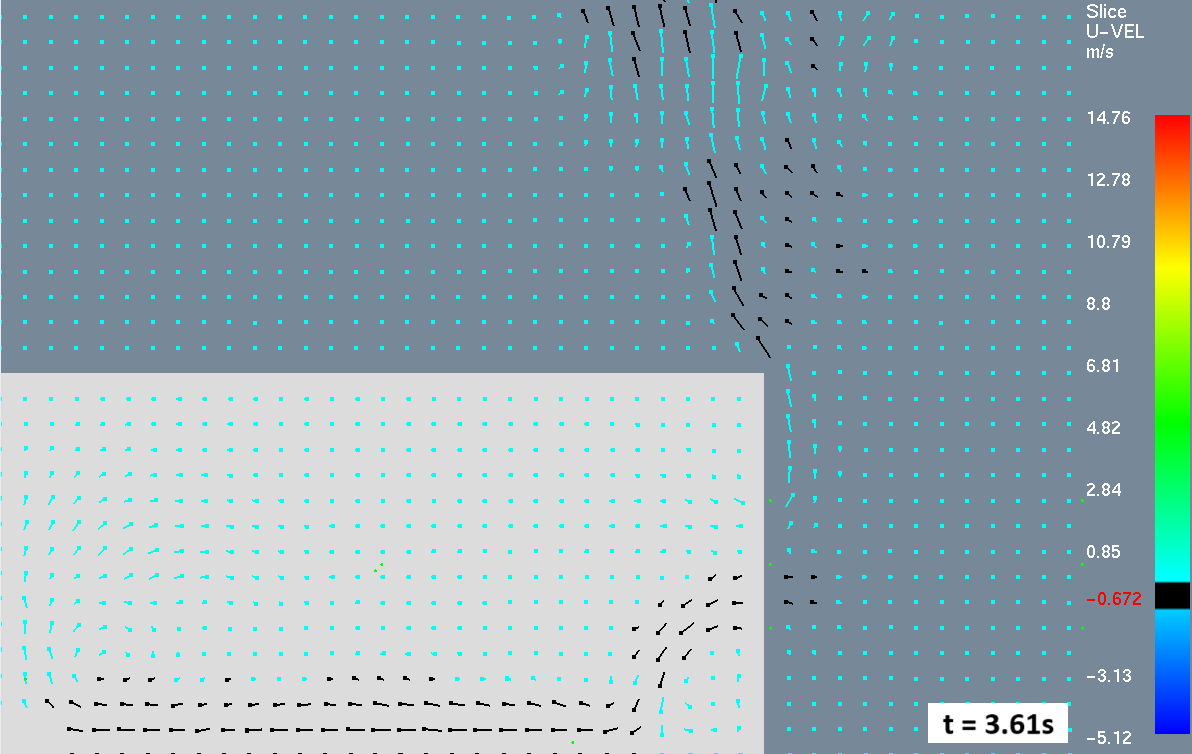}
	\caption{2-D velocity vector contour of model 2 under normal gravity conditions at the time of ignition on the central $y$-plane.}
	\label{inlet_vel_m2}
\end{figure*}

 \begin{figure*}[hbt!]  
	\centering
	\includegraphics[scale=0.55]{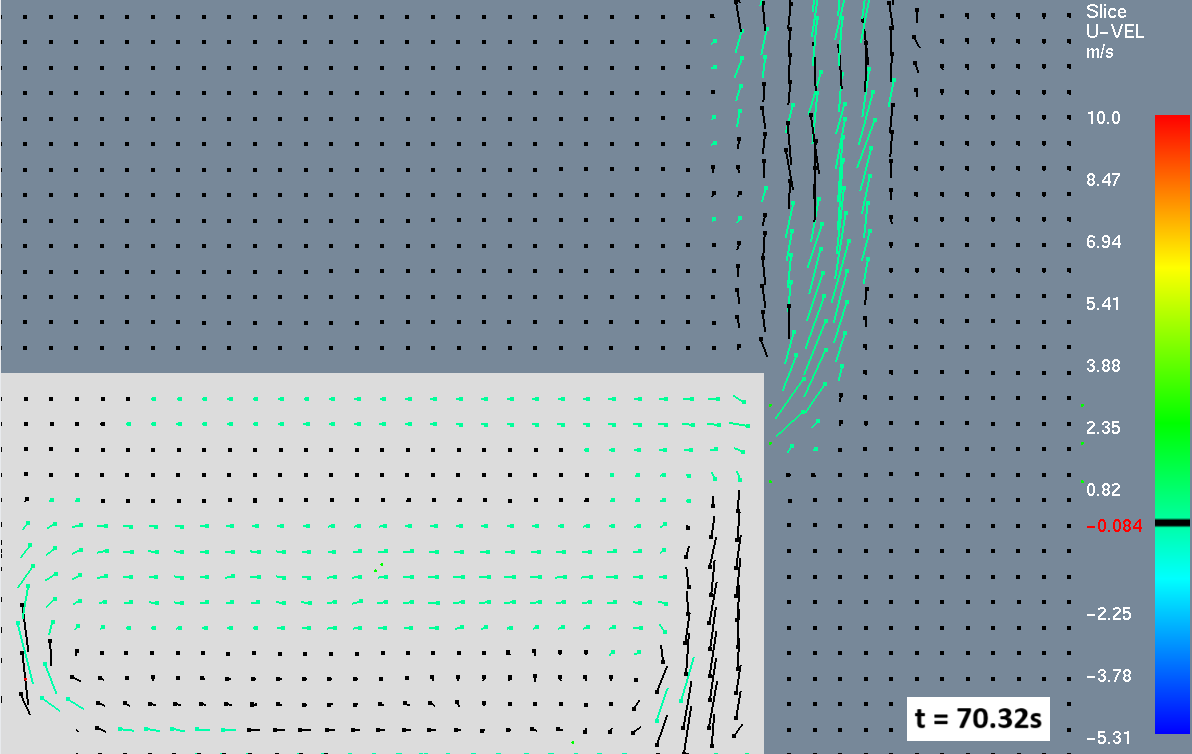}
	\caption{2-D velocity vector contour of model 3 under normal gravity conditions at the time of ignition on the central $y$-plane.}
	\label{inlet_vel_m3}

\vspace{0.5cm}

	\centering
	\includegraphics[scale=0.55]{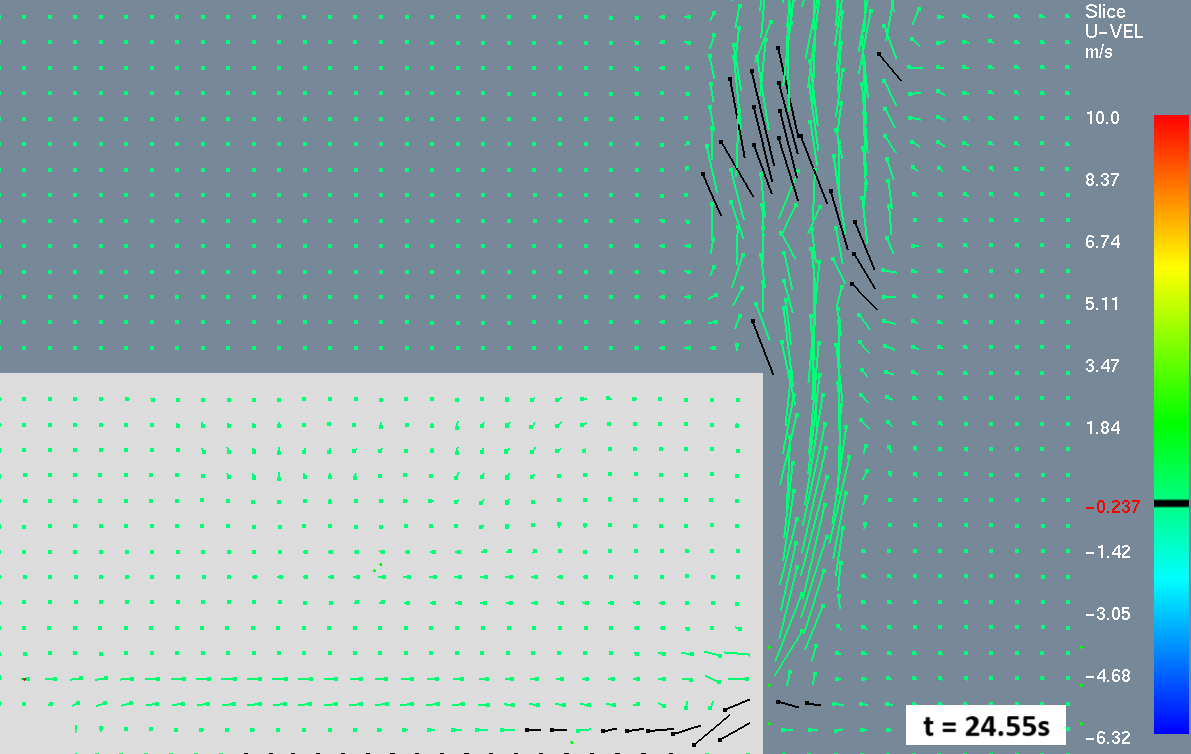}
	\caption{2-D velocity vector contour of model 4 under normal gravity conditions at the time of ignition on the central $y$-plane.}
	\label{inlet_vel_m4}
\end{figure*}

\begin{figure*}[!ht]
 \centering
 \begin{subfigure}[b]{0.48\textwidth}
     \centering
     \includegraphics[scale=0.26, trim={1cm 1cm 1cm 1cm},clip]{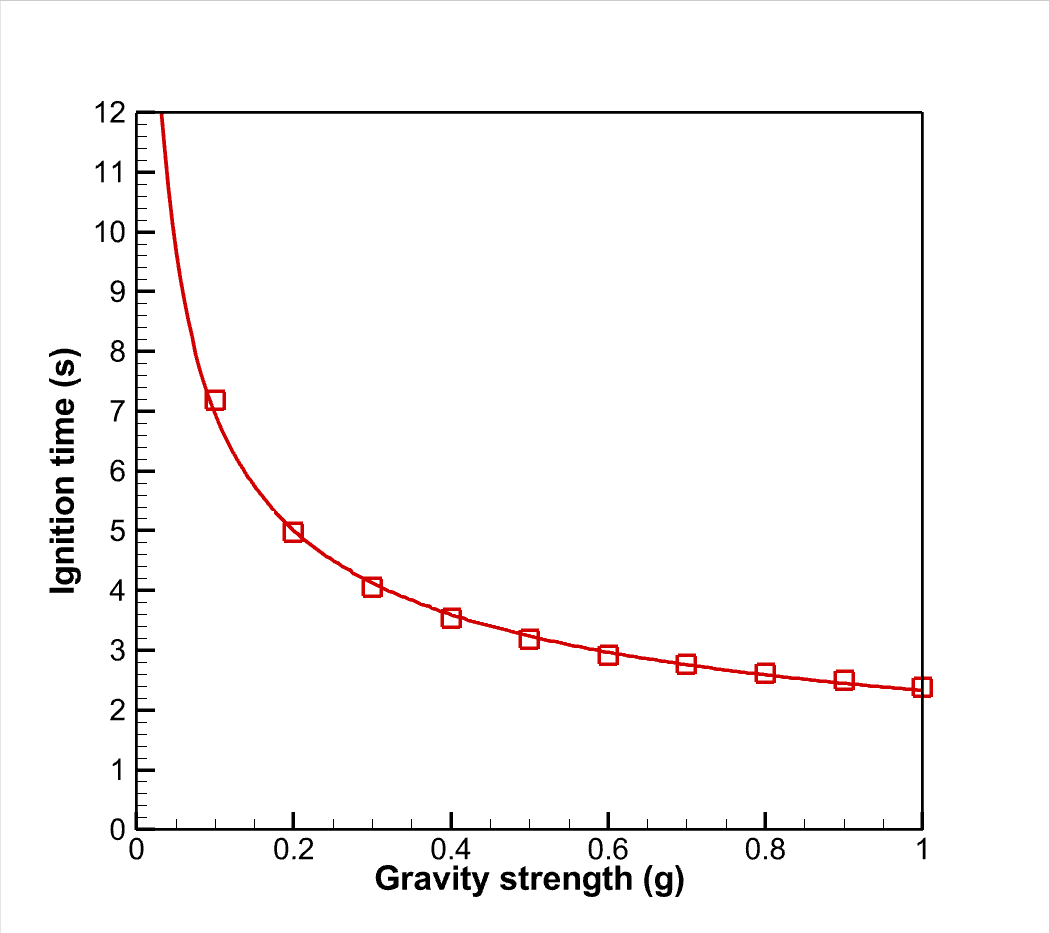}
     \caption{Model 1}
     \label{ign_model1}
 \end{subfigure}
 \hfill
 \begin{subfigure}[b]{0.48\textwidth}
     \centering
     \includegraphics[scale=0.26, trim={1cm 1cm 1cm 1cm},clip]{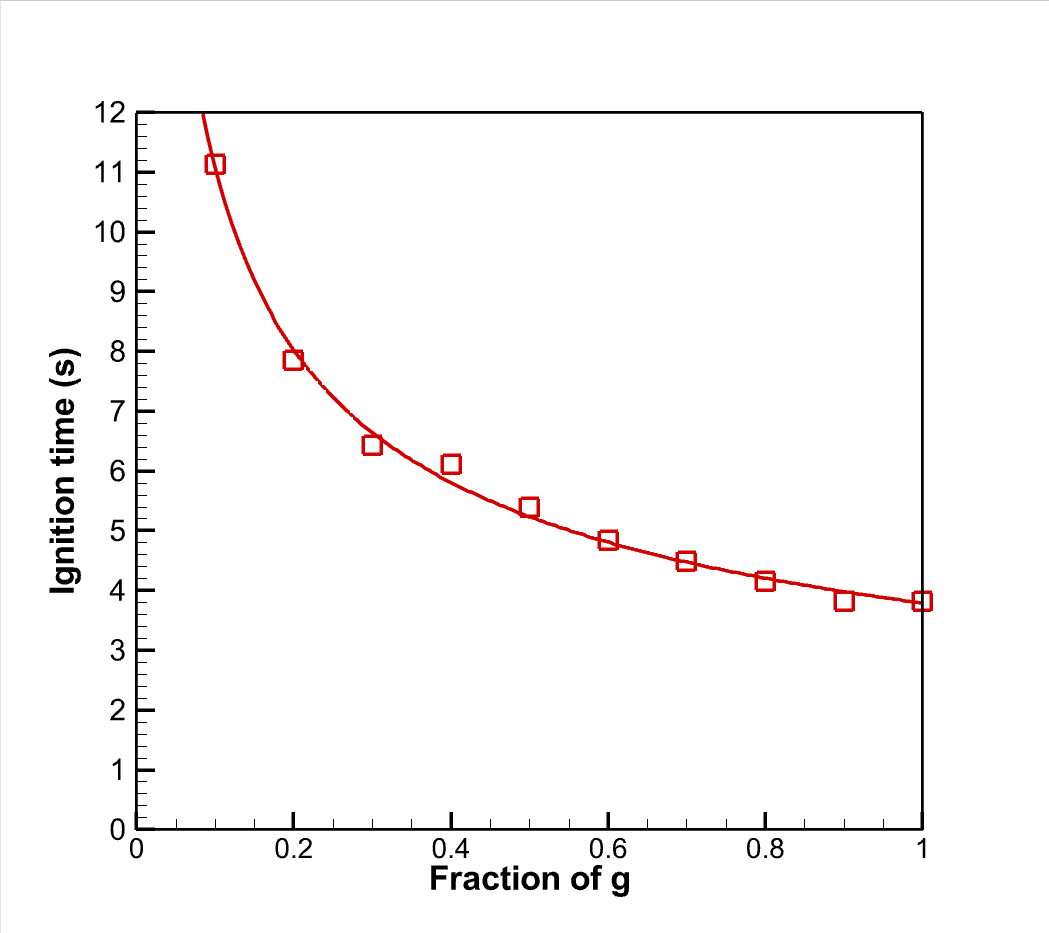}
     \caption{Model 2}
     \label{ign_model2}
 \end{subfigure}
 \vskip\baselineskip
 \begin{subfigure}[b]{0.48\textwidth}
     \centering
     \includegraphics[scale=0.26, trim={1cm 1cm 1cm 1cm},clip]{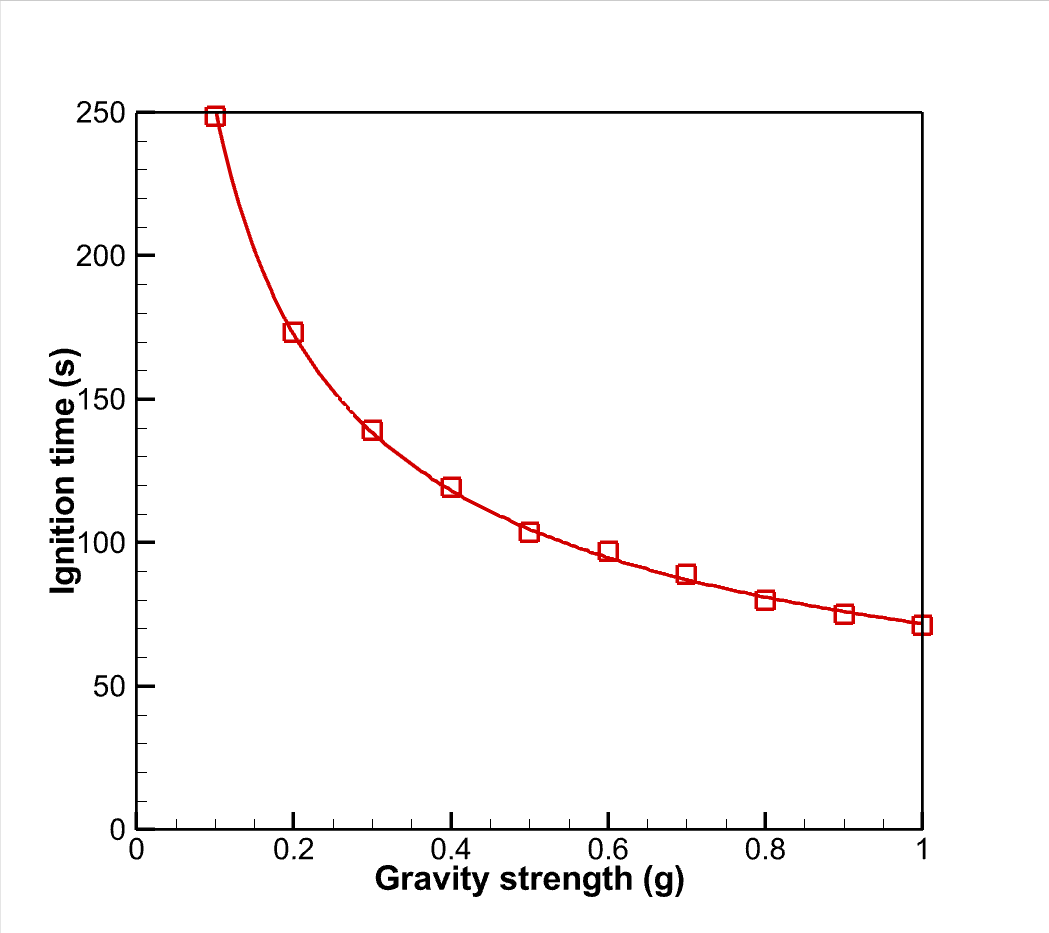}
     \caption{Model 3}
     \label{ign_model3}
 \end{subfigure}
\hfill
 \begin{subfigure}{0.48\textwidth}
     \centering
     \includegraphics[scale=0.26, trim={1cm 1cm 1cm 1cm},clip]{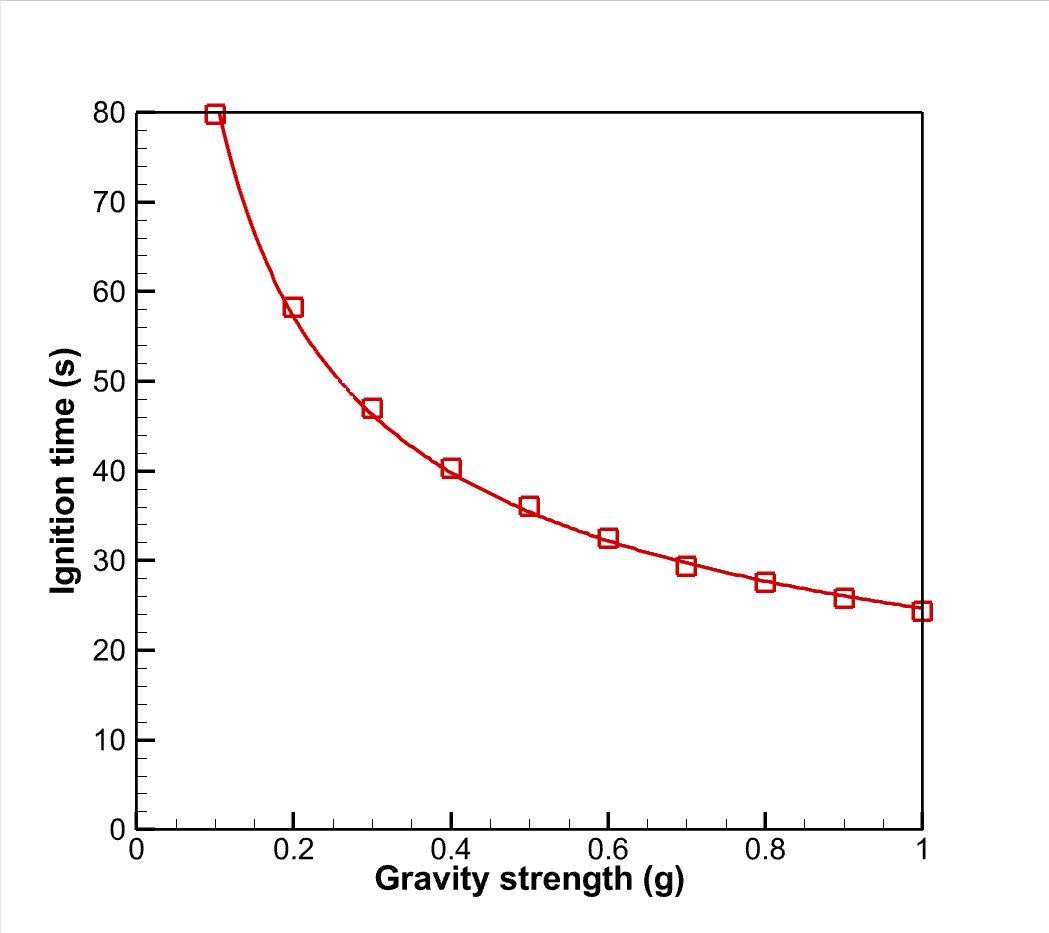}
     \caption{Model 4}
     \label{ign_model4}
 \end{subfigure}
 
 \caption{Ignition time vs. gravity strength for the four models. Red squares are data points, and the line is a power fit.}
 \label{igntime_plot}
\end{figure*}

Regarding models 3 and 4, the opening area is constricted to less than 20 \% of models 1 and 2. This is one of the reasons for the higher pressure. However, in model 4, which has the same opening area as model 3, there is a reduction in maximum pressure values compared to model 3 but significantly higher than models 1 and 2. 

Figures~\ref{inlet_vel_m1} through \ref{inlet_vel_m4} show the 2-D velocity vector contour under normal gravity conditions at the time of ignition in the central $y$-plane. The black section in the color bar shows the velocity within the range of that section. The velocity value shown in red corresponds to the point at the center of the black section. These images illustrate the flow of cold air coming in at that instant on that plane.

The velocity of incoming air is slightly lower for model 2 compared model 1. This is also reflected in a higher ignition delay time for model 2. The velocity is lowest for model 3 followed by model 4. Consequently, the ignition time is the largest for model 3. Therefore, the time for air entrainment from the opening is also the largest for model 3. Thus the peak pressures in model 3 are much higher than in model 4. The maximum pressure results obtained from this work are consistent with the observations made in Weng et al.~\cite{weng2005}, Ferraris et al.~\cite{Ferraris_2009}, and Myilsamy et al.~\cite{myilsamy_large_2019}. Despite these variations, these pressure values are negligible compared to the total thermodynamic pressure. This is because the combustion mode in backdraft is deflagration.

From the pressure results, two additional key observations can be made. First, the magnitude of the pressure rise remains relatively small compared to the absolute pressure. For models 1 and 2, the pressure rise is very small and, in the case of model 1, the noise associated with the velocity may have stronger effects than the actual trends associated with gravity. The highest pressure rise is as high as 2\% of the absolute pressure for model 3. A Less than 1\% rise is recorded for model 4.

Figure~\ref{igntime_plot} shows the ignition delay times as functions of the gravity magnitudes for the different models. For all cases, the ignition time decreases nonlinearly with the gravity strength. Model 1 is very similar to the work of Ashok and Echekki~\cite{ashok_numerical_2021} who noted similar results and similar trends. As discussed by Ashok and Echekki~\cite{ashok_numerical_2021}, the trend is expected because higher gravity strengths also result in higher speeds for the gravity currents. The trend is also consistent with the theory von Karman~\cite{von_karman_engineer_1940} and Benjamin~\cite{benjamin_gravity_1968} stated that the gravity current velocity is directly proportional to the square root of gravity strength. Considering the distance between the opening and the heat source a constant, we can also say that the ignition time is inversely proportional to the square root of gravity strength, i.e., $t_{\rm ign} \propto 1\left/\sqrt{g}\right.$. A decrease in gravity will reduce buoyancy thus the cold, denser air from outside takes a longer time to reach the farther end of the enclosure where the heat source is located.

\begin{figure*}[!ht]
 \centering
 \begin{subfigure}[b]{0.48\textwidth}
     \centering
     \includegraphics[scale=0.26, trim={1cm 1cm 1cm 1cm},clip]{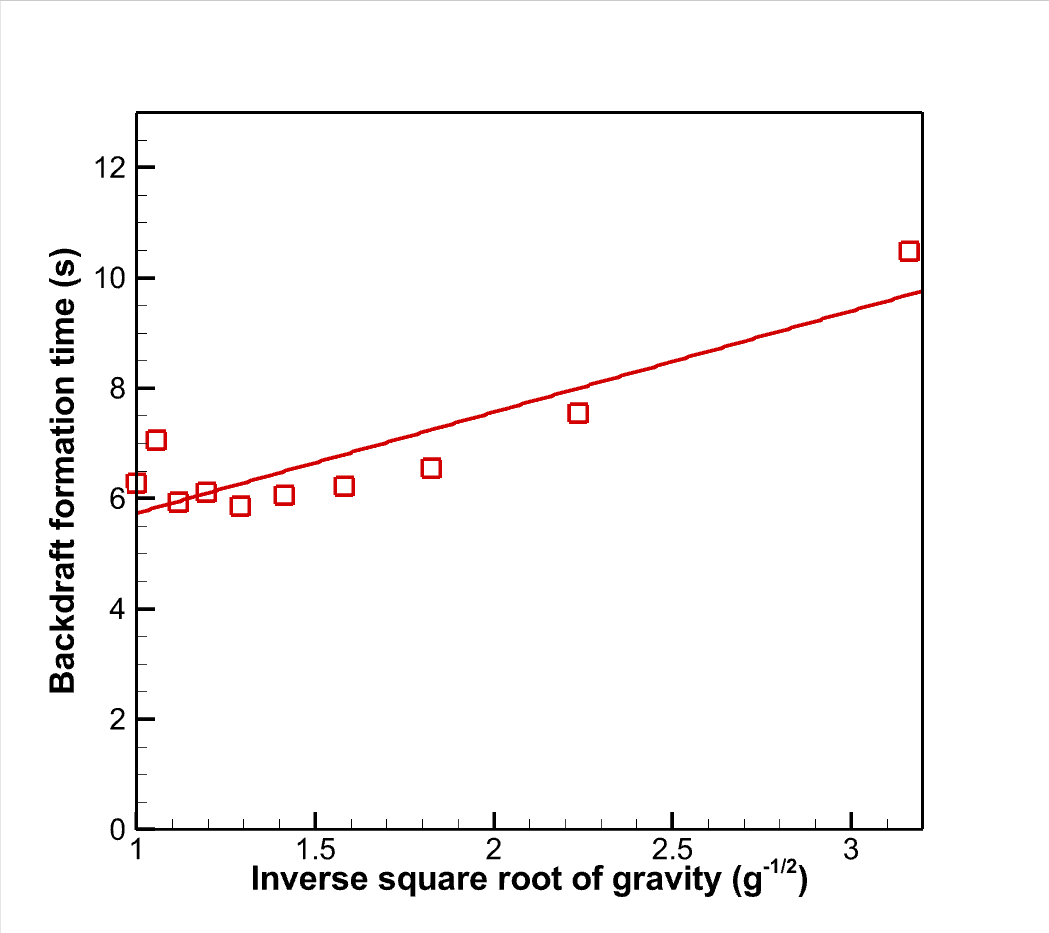}
     \caption{Model 1}
     \label{backdrafttime_model1}
 \end{subfigure}
 \hfill
 \begin{subfigure}[b]{0.48\textwidth}
     \centering
     \includegraphics[scale=0.26, trim={1cm 1cm 1cm 1cm},clip]{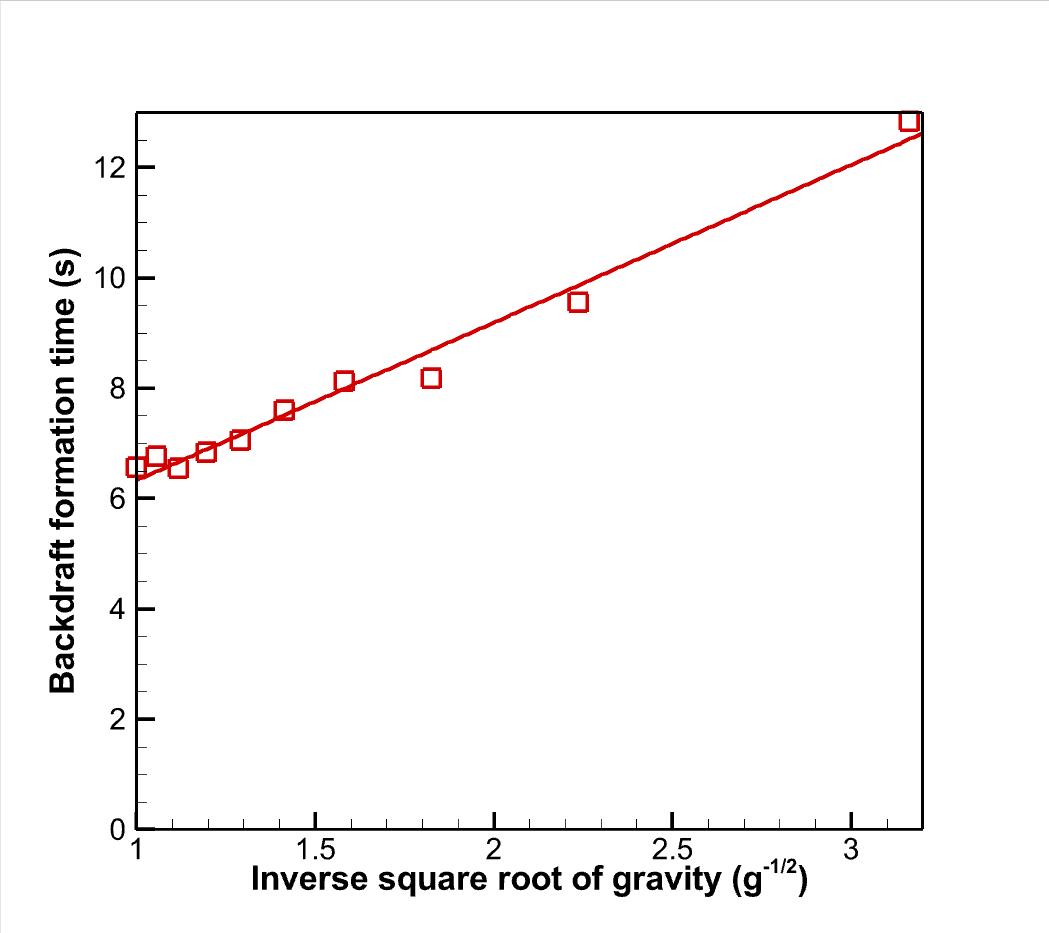}
     \caption{Model 2}
     \label{backdrafttime_model2}
 \end{subfigure}
 \vskip\baselineskip
 \begin{subfigure}[b]{0.48\textwidth}
     \centering
     \includegraphics[scale=0.26, trim={1cm 1cm 1cm 1cm},clip]{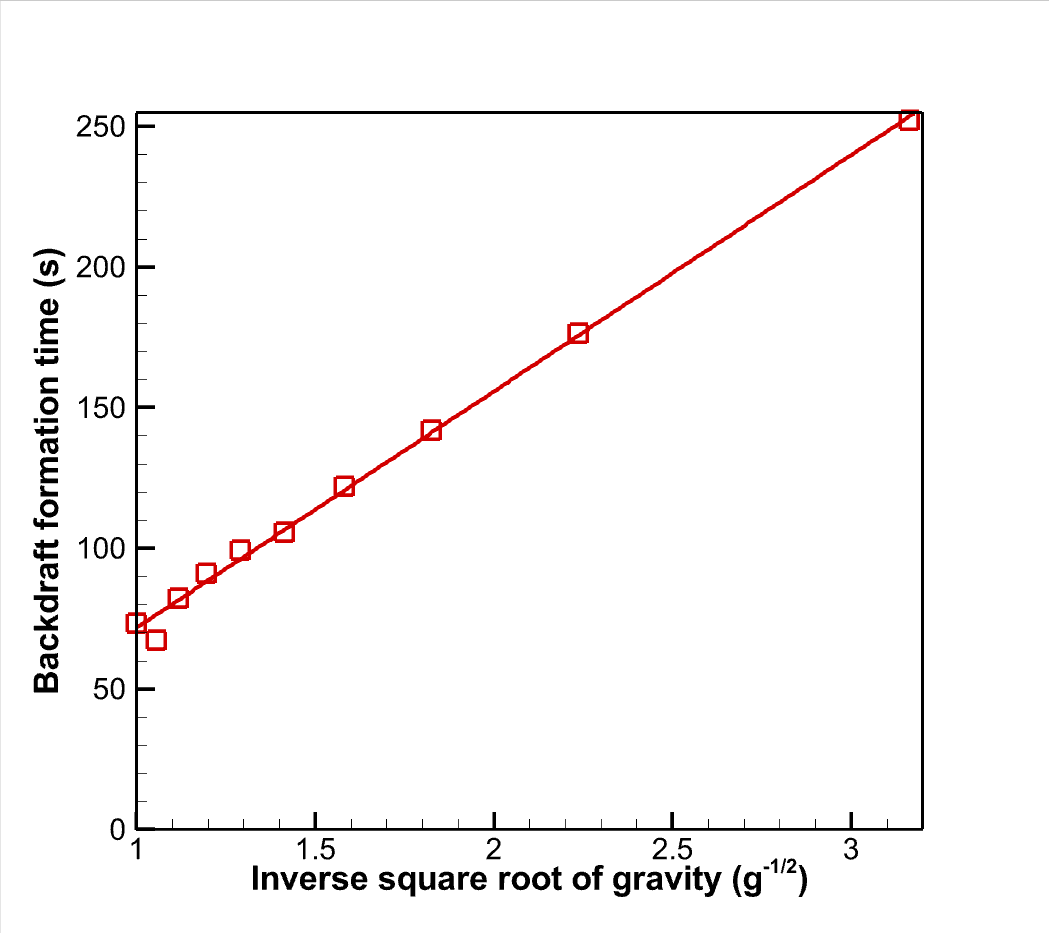}
     \caption{Model 3}
     \label{backdrafttime_model3}
 \end{subfigure}
\hfill
 \begin{subfigure}{0.48\textwidth}
     \centering
     \includegraphics[scale=0.26, trim={1cm 1cm 1cm 1cm},clip]{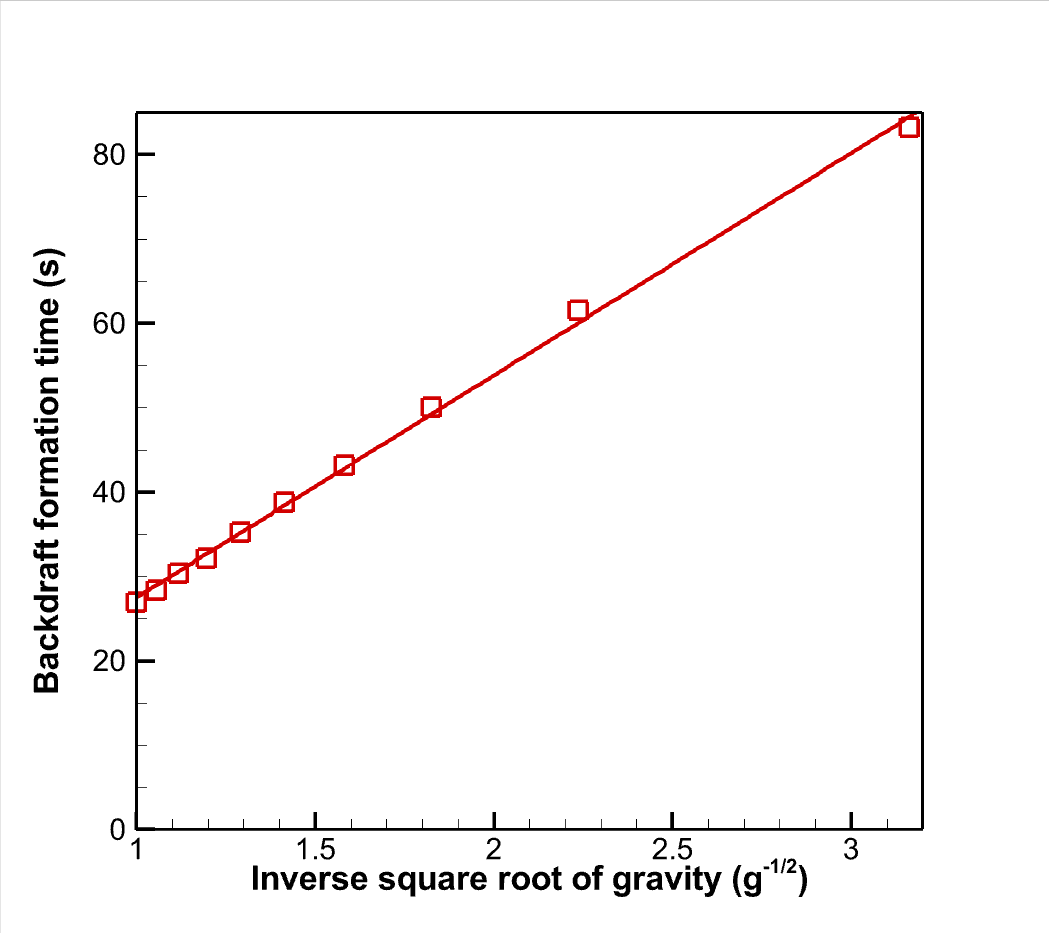}
     \caption{Model 4}
     \label{backdrafttime_model4}
 \end{subfigure}
 
 \caption{Backdraft formation time vs. inverse square root of gravity strength for the four models. Red squares are data points, and the line is a linear fit.}
 \label{backdrafttime_plot}
\end{figure*}

The relation between ignition time $t_{\rm ign}$ and gravity strength holds well for all the cases studied in this work. Yet, the magnitudes of the ignition times or the proportionality constants are different. Models 1 and 2 exhibit similar values and ranges. However, models 3 and 4 exhibit a much higher delay, with model 3 exhibiting more than an order of magnitude increase in ignition time for the same enclosure geometry. 

Another parameter of interest is the backdraft formation time. This is the time at which the deflagration exits the enclosure. This time is plotted against the inverse square root of gravity strength for the four models as shown in Figure~\ref{backdrafttime_plot}. The backdraft formation time is plotted vs. the inverse of the square root of the gravity constant to see if the scaling relation for the ignition time based on von Karman~\cite{von_karman_engineer_1940} and Benjamin~\cite{benjamin_gravity_1968} holds. The figure clearly shows that this relation holds for all models, suggesting that the gravity current is the primary mechanism for both the onset of ignition and backdraft. 

\begin{figure*} 
	\centering
	\includegraphics[scale=.65]{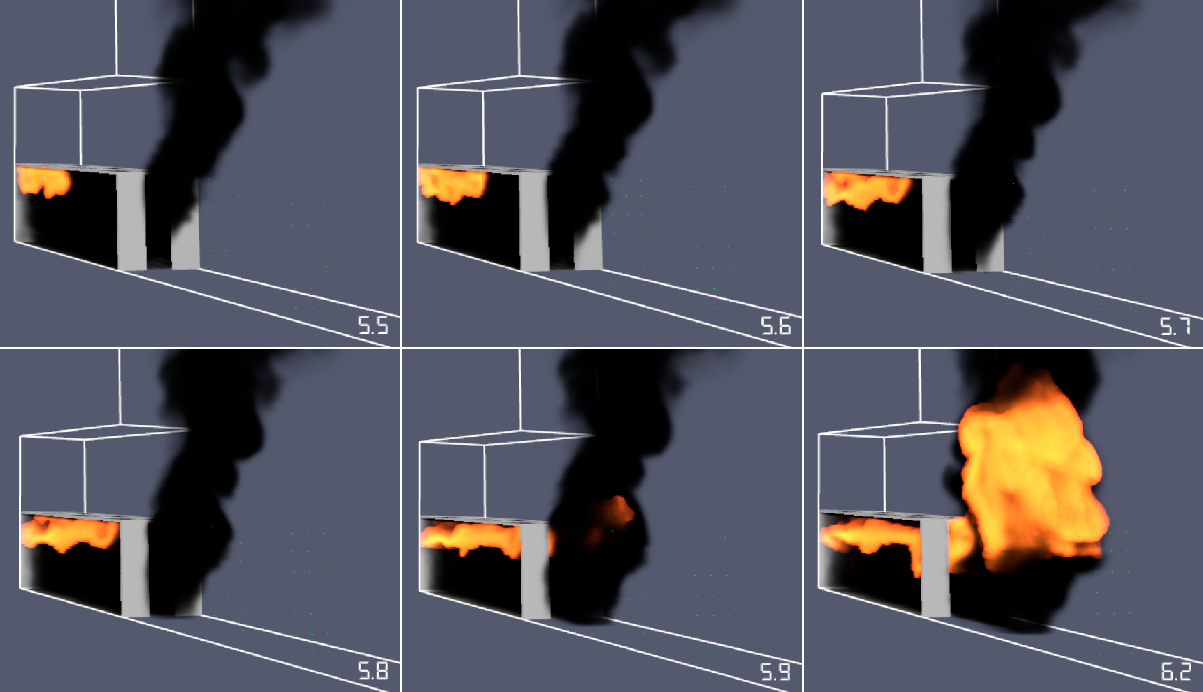}
	\caption{3-D rendering of soot and fire of model 1 under normal gravity conditions.}
	\label{3d_soot_model1}

 \vspace{0.5cm}

	\centering
	\includegraphics[scale=.65]{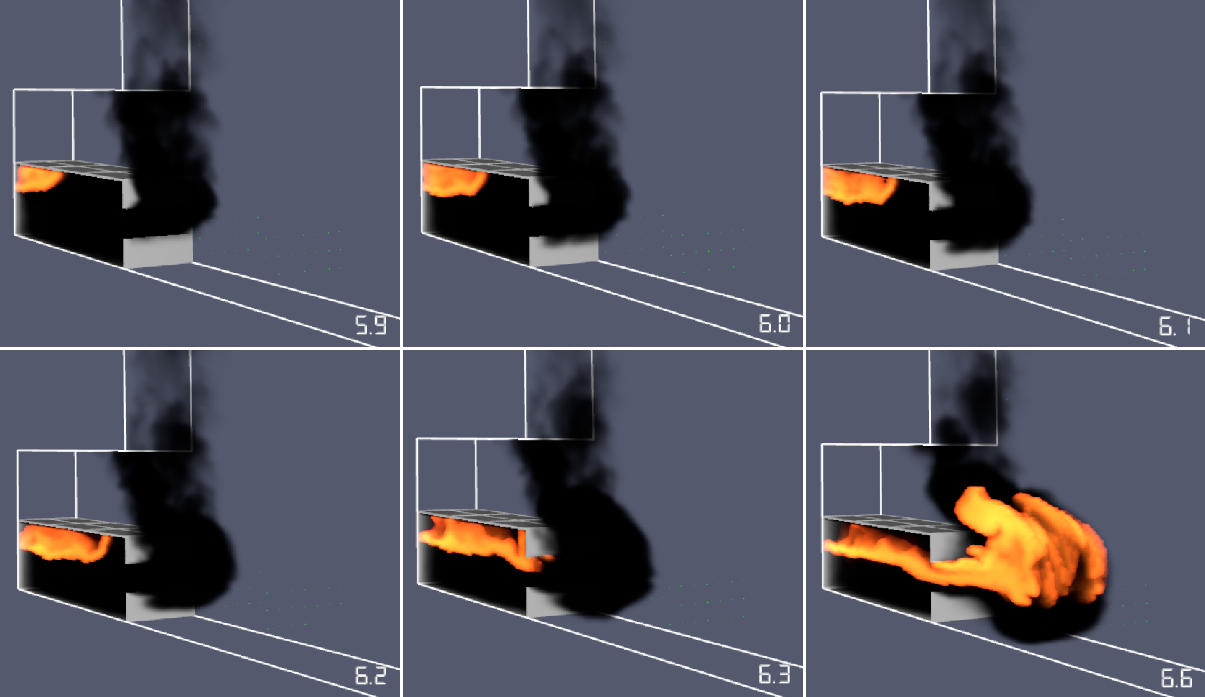}
	\caption{3-D rendering of soot and fire of model 2 under normal gravity conditions.}
	\label{3d_soot_model2}
\end{figure*}

\begin{figure*}
	\centering
	\includegraphics[scale=.65]{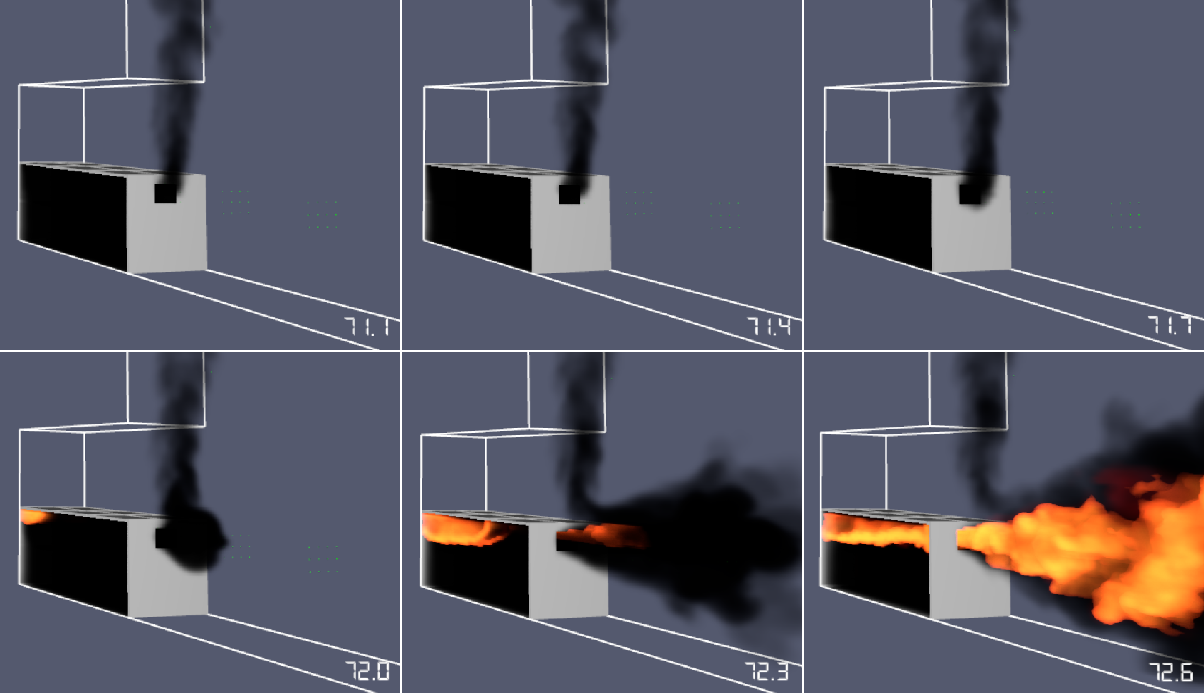}
	\caption{3-D rendering of soot and fire of model 3 under normal gravity conditions.}
	\label{3d_soot_model3}

\vspace{0.5cm}

	\centering
	\includegraphics[scale=.65]{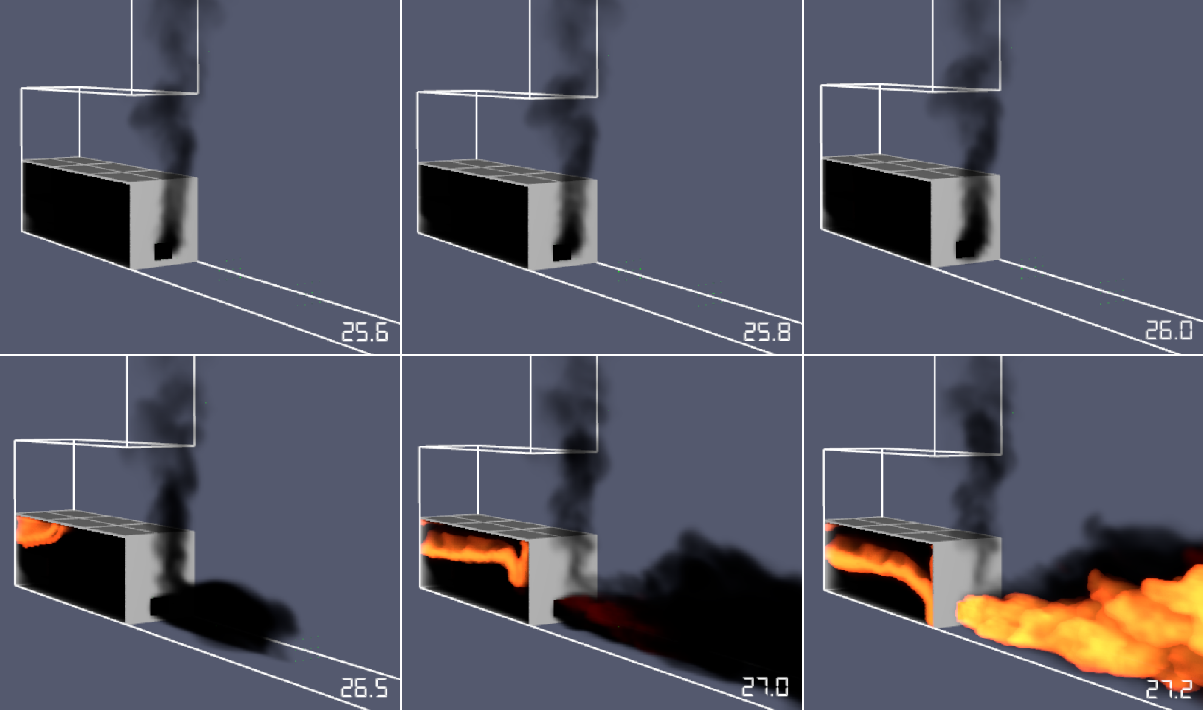}
	\caption{3-D rendering of soot and fire of model 4 under normal gravity conditions.}
	\label{3d_soot_model4}
\end{figure*}

Next, we assess if the smoke release can serve as a precursor for the onset of backdraft. Figures~\ref{3d_soot_model1} to \ref{3d_soot_model4} show the three-dimensional rendering of soot and fire for the four models under normal gravity conditions. The time of each frame is shown at the bottom left section of each snippet. This time has been selected such that the increase in soot and deflagration propagation is visible. 

In model 1, the soot starts to increase at approximately 5.7 s and the deflagration wave reaches the exit at 5.8 s. Thus, there is approximately 0.1 s available for taking precautions. Once the hatch is opened, the existing smoke inside the enclosure starts coming out almost at a steady rate. However, as the deflagration exits the opening and indicates the onset of backdraft, the smoke volume increases. An arbitrary soot mass fraction value of nearly 0.01 kg/kg is considered above which a perceivable increase in the soot mass fraction occurs for models 1 and 2. This soot mass fraction value for models 3 and 4 is close to 0.005 kg/kg.

In model 2, the soot starts to increase at around 6 s and deflagration reaches the exit at 6.3 s, thus giving a window of 0.3 s to take precautions. In model 3, the soot starts to increase at around 71.7 s and deflagration reaches the exit at 72.3 s. Here, we have around 0.6 s to take precautions. Finally, for model 4, the soot can be seen increasing at approximately 26 s and deflagration exits the enclosure at 27s thus giving one second to take precautions. This analysis gives an estimate of the amount of time available for a person to take necessary precautions to avoid the effects of the deflagration wave based on observing a rise in smoke coming from the enclosure. 

The visual cue or the perception of an increase in the amount of smoke may be subjective. It also depends on the position of the soot mass fraction sensor at the exit plane of the enclosure as well as the combustion chemistry and the initial conditions. Yet, the results suggest that soot can be used as a precursor for the onset of deflagration even if the warning time may be considered short.

\begin{figure}[hbt!]  
	\centering
	\includegraphics[scale=0.26, trim={1cm 1cm 1cm 1cm},clip]{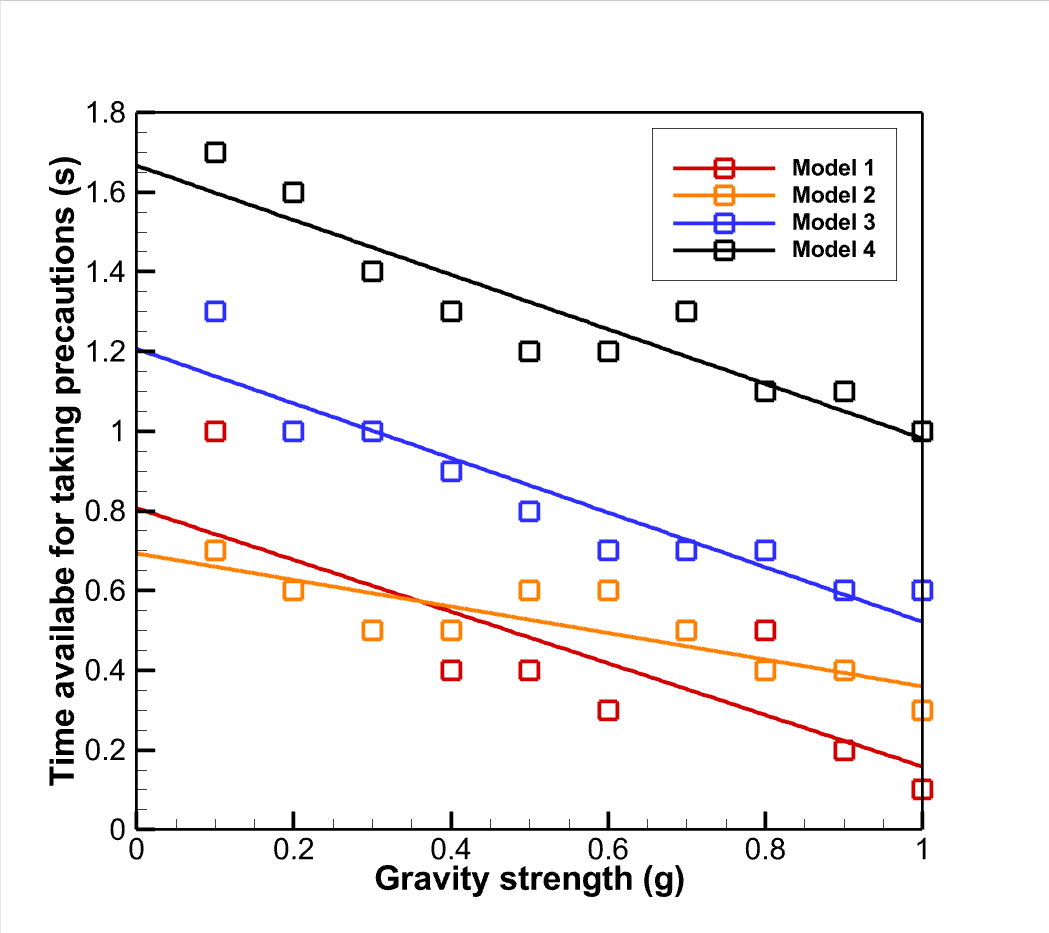}
	\caption{Time available for taking precautions vs. gravity strength for the four models. Colored squares are data points, and the lines are linear fit.}
	\label{time_precautions}
\end{figure}

Figure~\ref{time_precautions} shows the trend for the variation of the time available for taking precautions with gravity strength. This time decreases with the gravity constant for all the models. The maximum time is available for model 4 followed by model 3. Models 1 and 2 have similar time ranges. The trends among the different openings can be related to the reported values of the maximum pressure shown in Figure~\ref{maxP}. The higher the pressure difference between the compartments and the surroundings, the higher the rate of the expected purge of smoke from the enclosure. In contrast, the onset of backdraft may be related to this pressure difference and the rate of oxidation of the fuel in the enclosure. Regardless, we have just sufficient time to steer away from the path of the deflagration wave.

\begin{figure*}
 \centering
 \begin{subfigure}[b]{0.48\textwidth}
     \centering
     \includegraphics[scale=0.26, trim={1cm 1cm 1cm 1cm},clip]{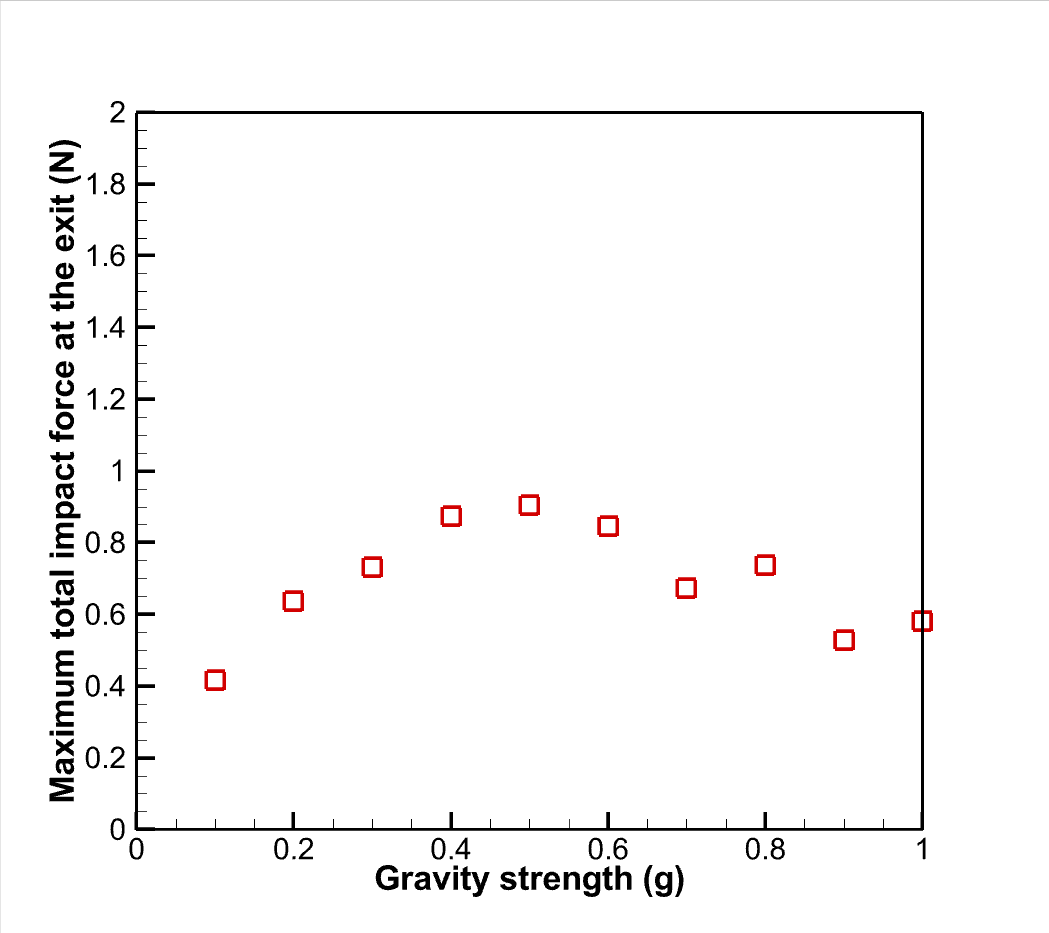}
     \caption{Model 1}
     \label{impact_model1}
 \end{subfigure}
 \hfill
 \begin{subfigure}[b]{0.48\textwidth}
     \centering
     \includegraphics[scale=0.26, trim={1cm 1cm 1cm 1cm},clip]{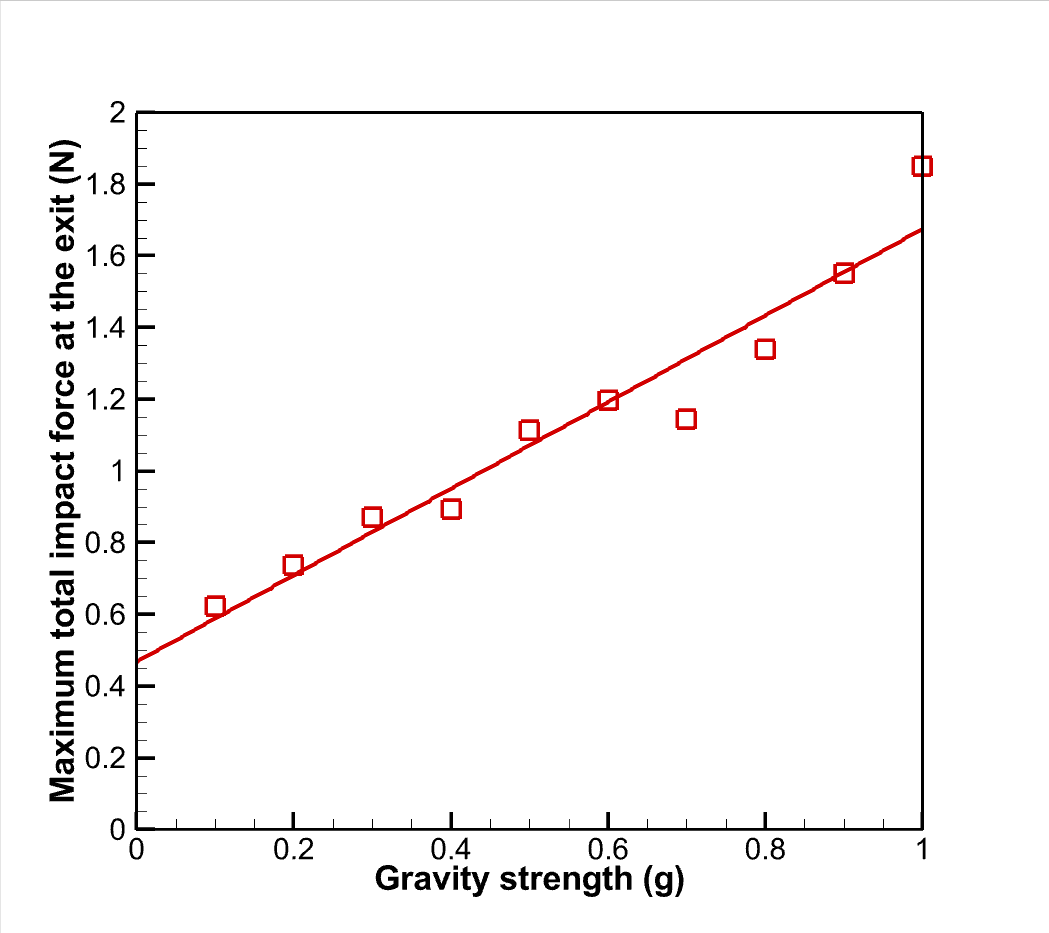}
     \caption{Model 2}
     \label{impact_model2}
 \end{subfigure}
 \vskip\baselineskip
 \begin{subfigure}[b]{0.48\textwidth}
     \centering
     \includegraphics[scale=0.26, trim={1cm 1cm 1cm 1cm},clip]{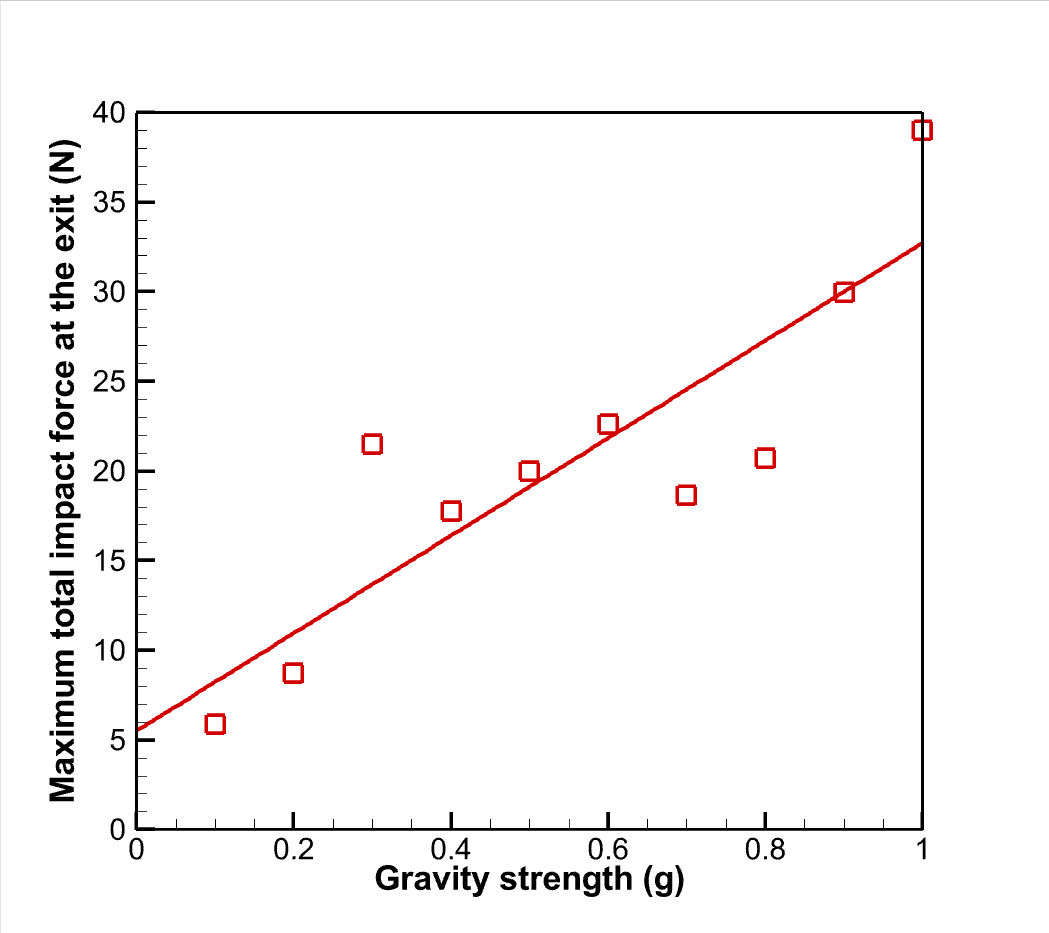}
     \caption{Model 3}
     \label{impact_model3}
 \end{subfigure}
\hfill
 \begin{subfigure}{0.48\textwidth}
     \centering
     \includegraphics[scale=0.26, trim={1cm 1cm 1cm 1cm},clip]{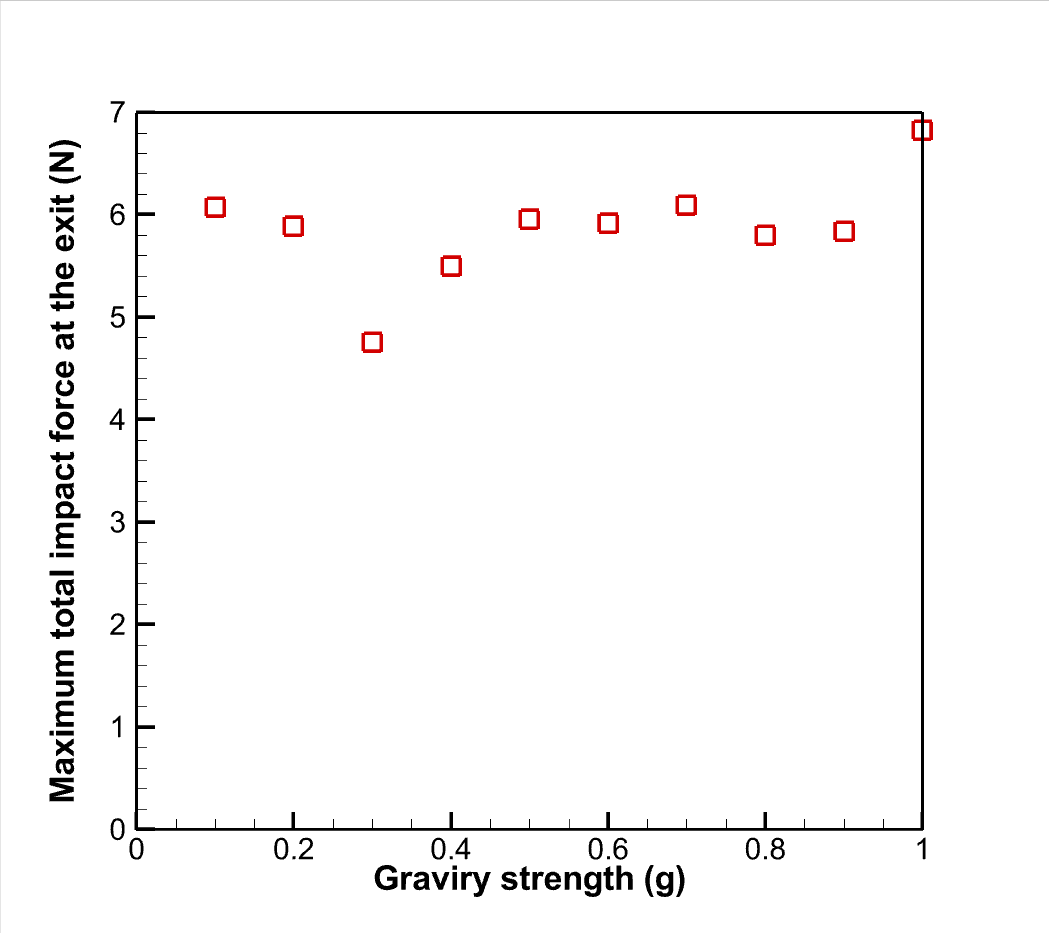}
     \caption{Model 4}
     \label{impact_model4}
 \end{subfigure}

 \caption{Maximum impact force at the exit vs. gravity strength for the four models. Red squares are data points, and the line is a linear fit.}
 \label{Momentum_flow_main} 
\end{figure*}

The final objective of this work is to analyze the effect of the backdraft in the form of impact force and thermal exposure based on Eqs.~(\ref{impact_eq}) and (\ref{heat_flow}). The variation of maximum impact force at the exit is plotted against the gravity strength for all four models as shown in Figure~\ref{Momentum_flow_main}. 

It is important to note that the trends of the impact force are closely related to the trends of the maximum pressure, as shown in Figure~\ref{maxP}. Both models 1 and 2 with similar opening areas and similar magnitudes of the maximum pressure also exhibit similar magnitudes of the maximum impact force. Model 3, which exhibits the highest maximum pressure, also exhibits the highest impact forces, which are an order of magnitude higher than the values for models 1 and 2. Also, note that there are no discernible trends of the impact force as a function of the gravity constant for models 1 and 4, as also observed for the maximum pressure.

\begin{figure*}
	\centering
	\includegraphics[scale=.50]{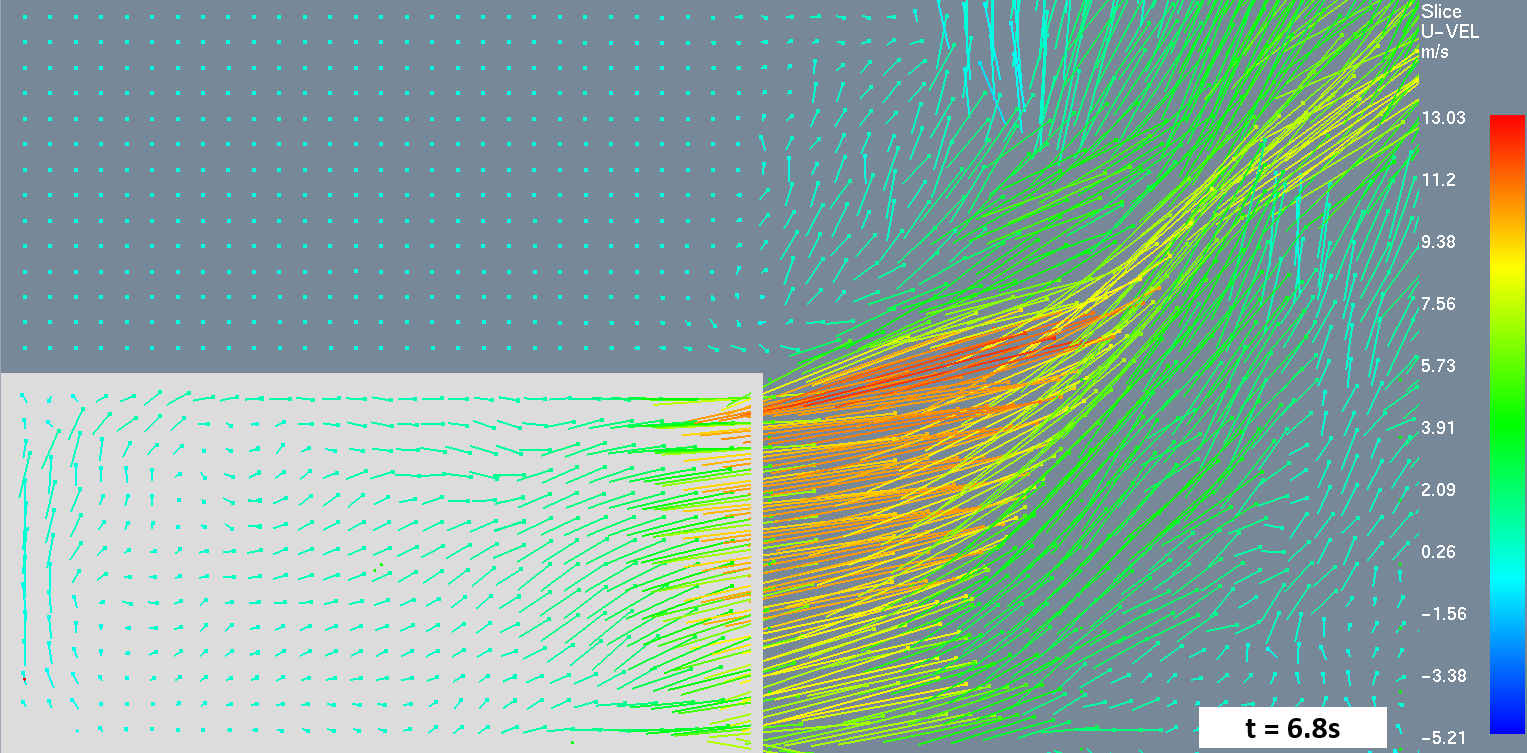}
	\caption{2-D velocity vector contour of model 1 under normal gravity conditions at the time when maximum impact force occurs at the exit on the central $y$-plane.}
	\label{vel_contour_maxF1}

\vspace{0.5cm}

	\centering
	\includegraphics[scale=.50]{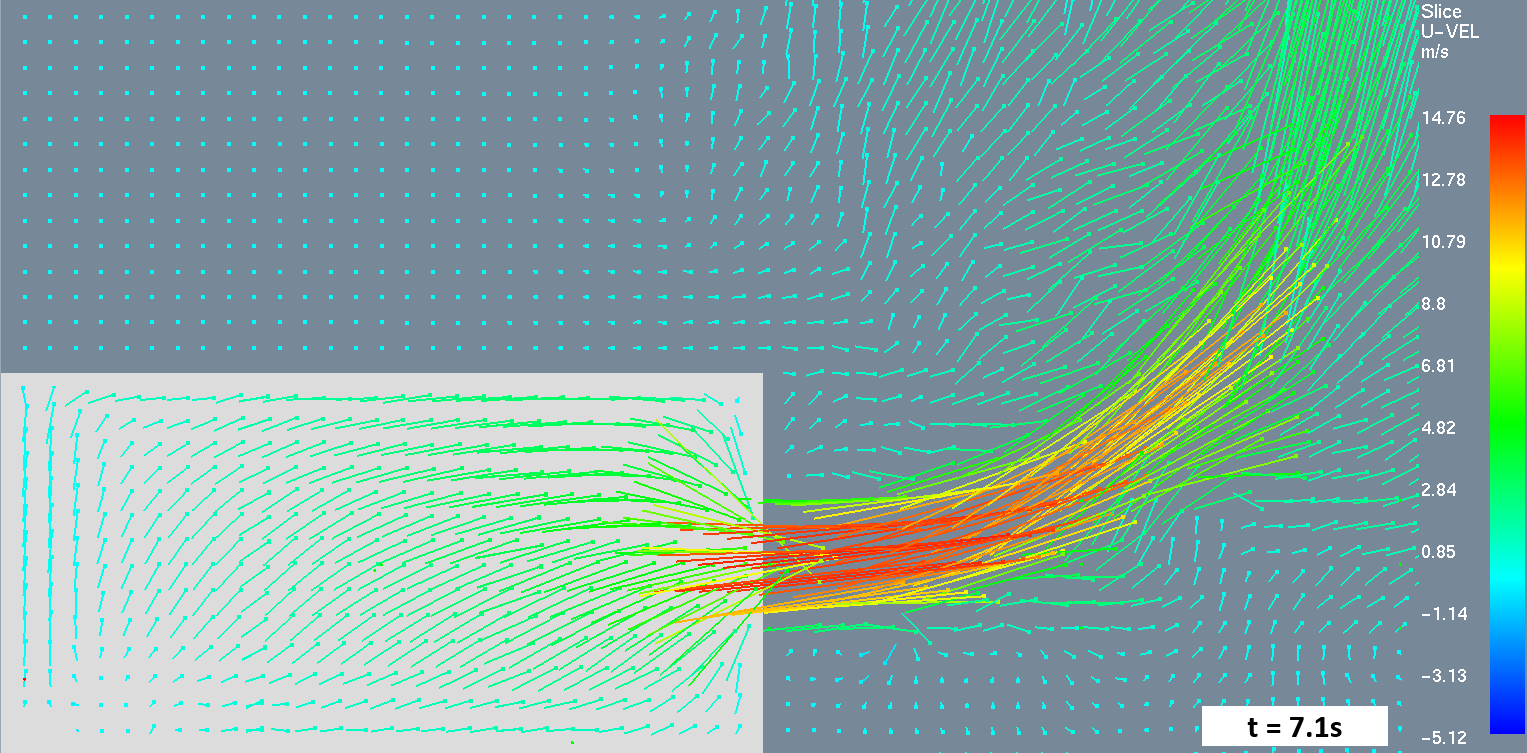}
	\caption{2-D velocity vector contour of model 2 under normal gravity conditions at the time when maximum impact force occurs at the exit on the central $y$-plane.}
	\label{vel_contour_maxF2}
\end{figure*}

\begin{figure*}
	\centering
	\includegraphics[scale=.50]{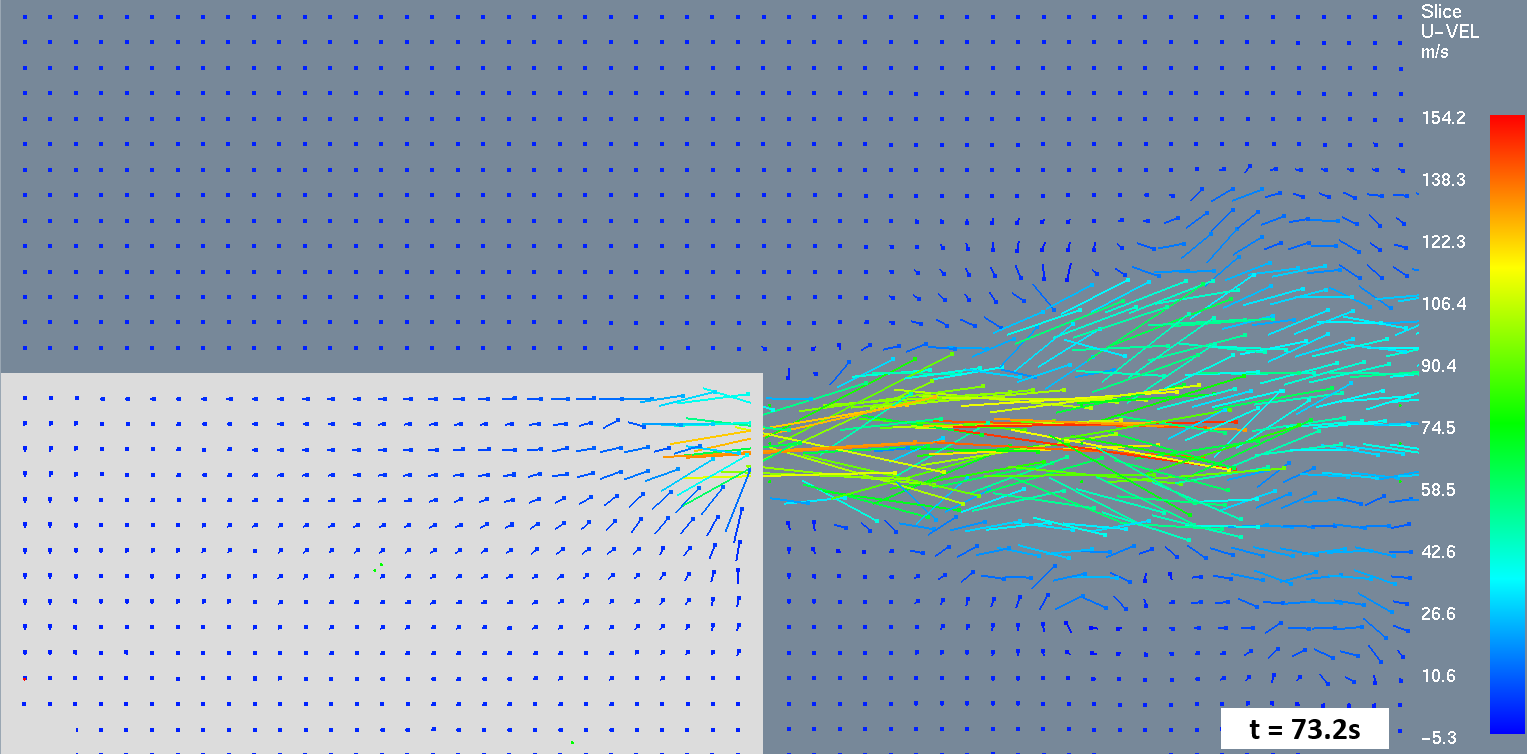}
	\caption{2-D velocity vector contour of model 3 under normal gravity conditions at the time when maximum impact force occurs at the exit on the central $y$-plane.}
	\label{vel_contour_maxF3}

\vspace{0.5cm}

	\centering
	\includegraphics[scale=.50]{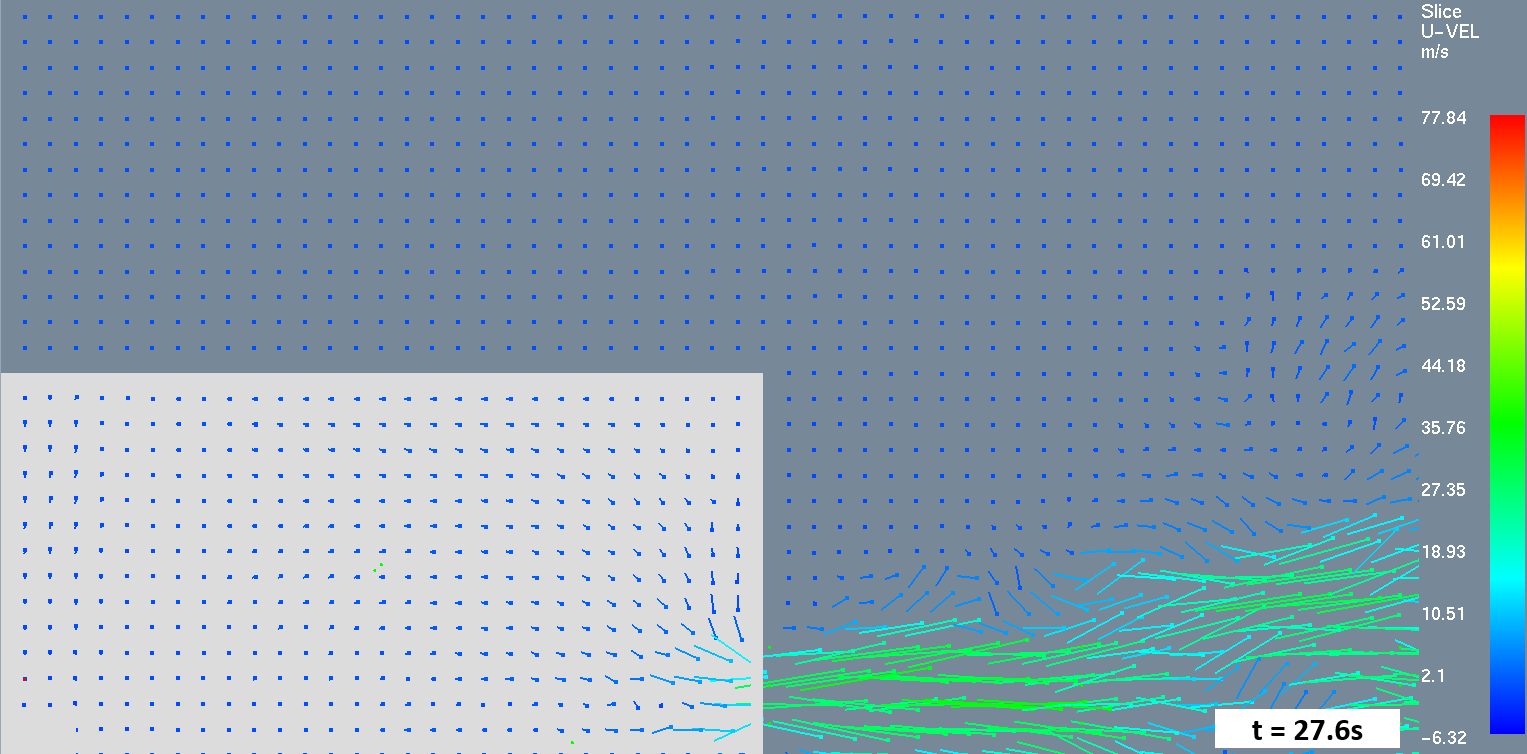}
	\caption{2-D velocity vector contour of model 4 under normal gravity conditions at the time when maximum impact force occurs at the exit on the central $y$-plane.}
	\label{vel_contour_maxF4}
\end{figure*}

Figures ~\ref{vel_contour_maxF1} to \ref{vel_contour_maxF4} show the 2-D velocity vector contour of the four models under normal gravity conditions at the time when maximum impact force occurs at the exit. It can be seen that the maximum $u$-velocity is the highest for model 3 followed by models 4, 2, and 1. According to equation~\ref{impact_eq}, the maximum impact force depends on $u^2$. This explains the variation in the maximum impact forces between the four models. It has to be noted that the impact forces generated from the backdraft are not strong enough to cause any injury to a person standing at the exit of the enclosure.

\begin{figure*}
 \centering
 \begin{subfigure}[b]{0.48\textwidth}
     \centering
     \includegraphics[scale=0.26, trim={1cm 1cm 1cm 1cm},clip]{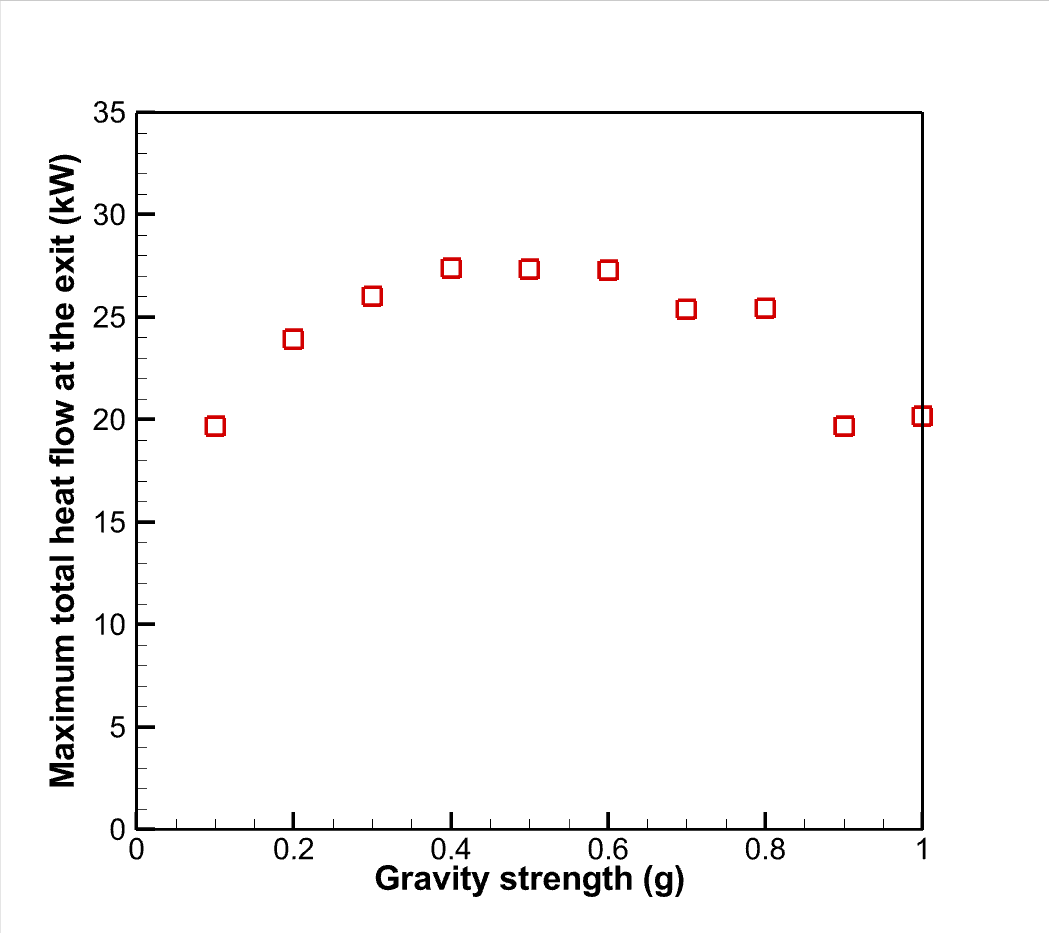}
     \caption{Model 1}
     \label{HT_model1}
 \end{subfigure}
 \hfill
 \begin{subfigure}[b]{0.48\textwidth}
     \centering
     \includegraphics[scale=0.26, trim={1cm 1cm 1cm 1cm},clip]{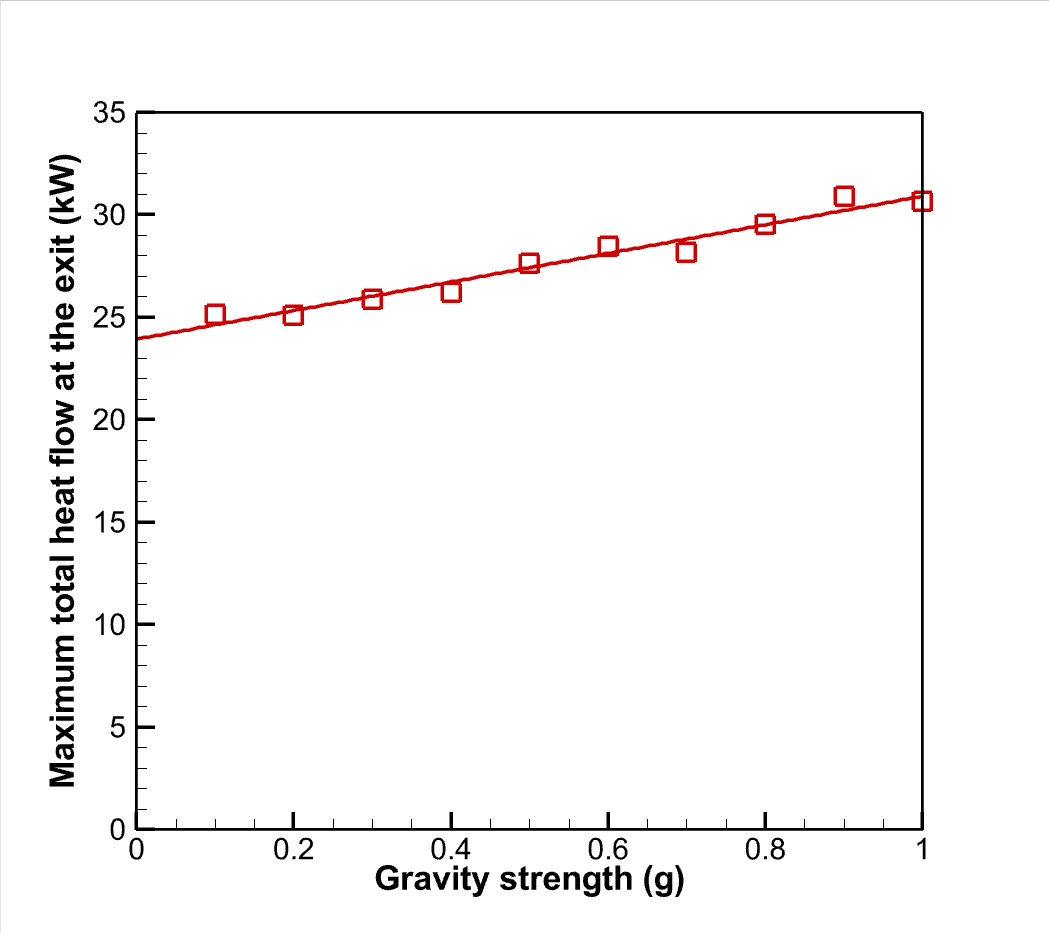}
     \caption{Model 2}
     \label{HT_model2}
 \end{subfigure}
 \vskip\baselineskip
 \begin{subfigure}[b]{0.48\textwidth}
     \centering
     \includegraphics[scale=0.26, trim={1cm 1cm 1cm 1cm},clip]{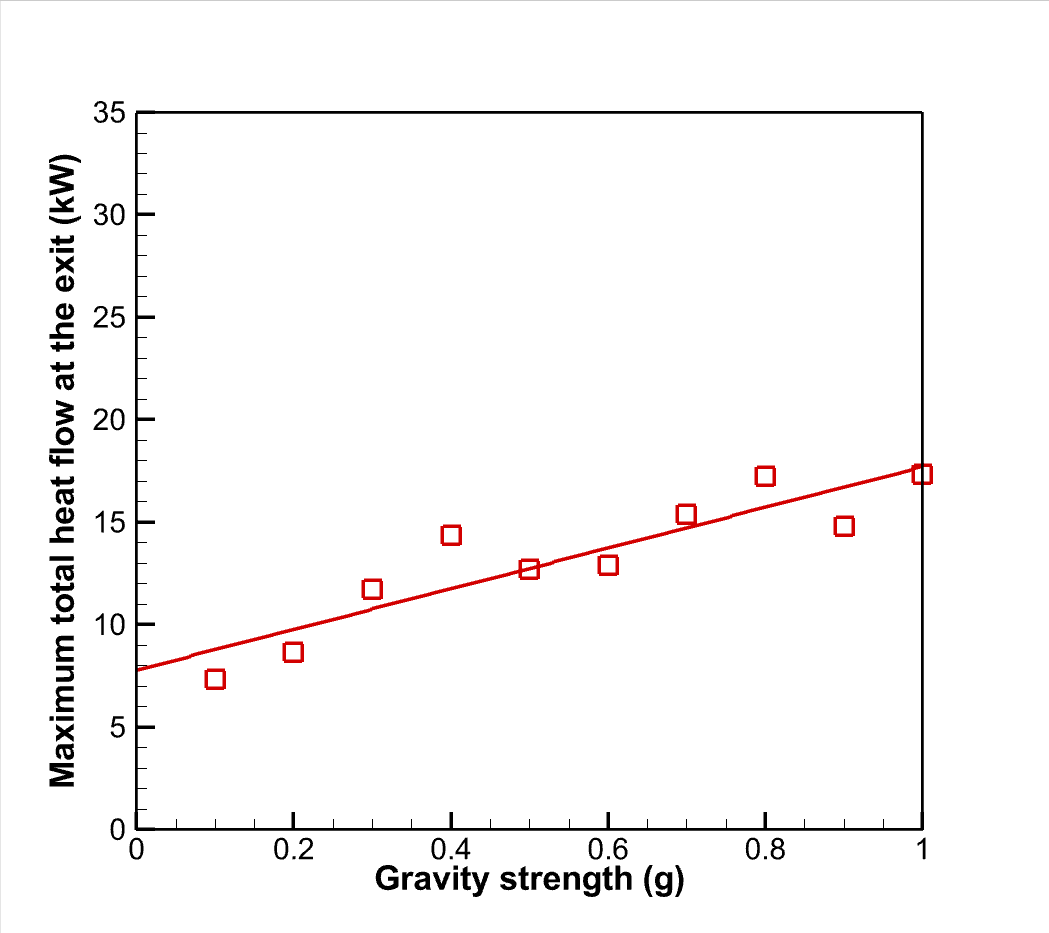}
     \caption{Model 3}
     \label{HT_model3}
 \end{subfigure}
\hfill
 \begin{subfigure}{0.48\textwidth}
     \centering
     \includegraphics[scale=0.26, trim={1cm 1cm 1cm 1cm},clip]{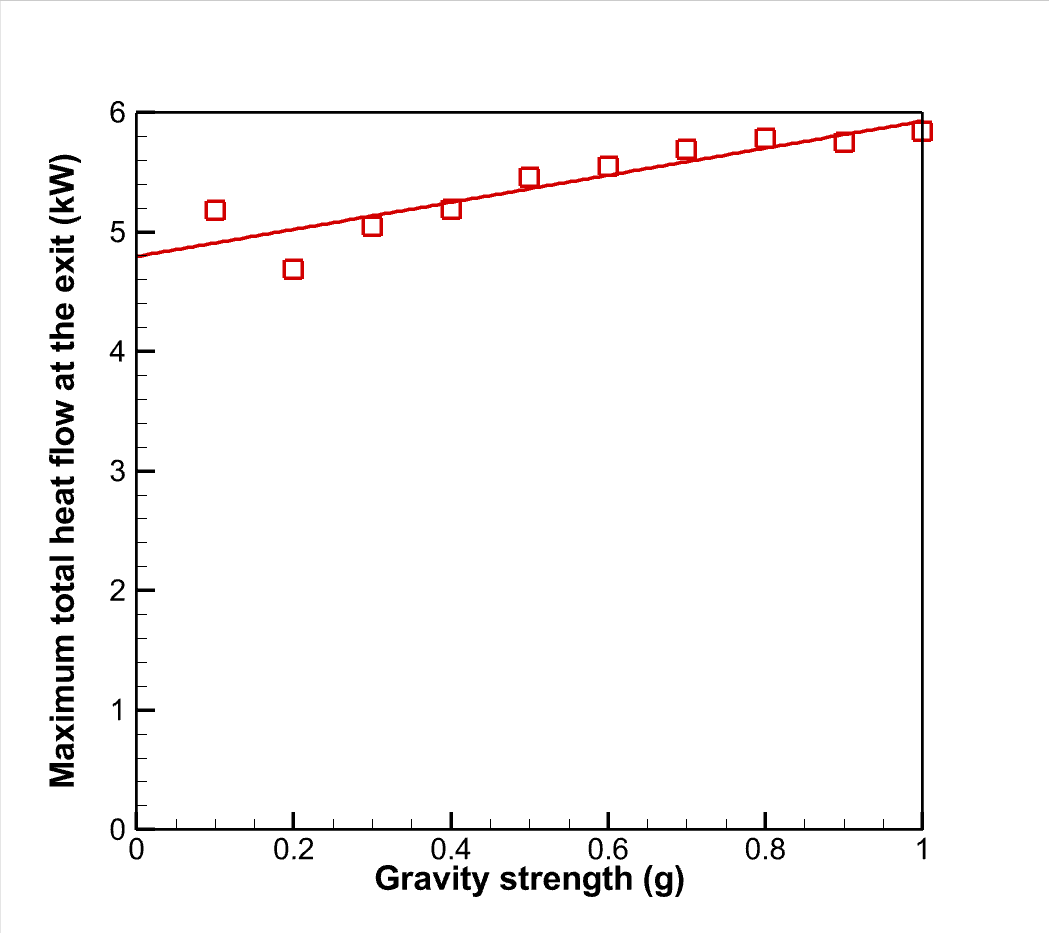}
     \caption{Model 4}
     \label{HT_model4}
 \end{subfigure}

 \caption{Maximum thermal exposure at the exit vs. gravity strength for the four models. Red squares are data points, and the line is a linear fit.}
 \label{Heat_flow_main}
\end{figure*}

Figure~\ref{Heat_flow_main} shows the maximum thermal heating variation through the exit with gravity strength for the four models based on Eq.~(\ref{heat_flow}). These plots also follow similar trends as the maximum impact force plots. Model 1 has no clear trend for the maximum thermal exposure. The maximum thermal exposure for models 2, 3, and 4 decreases with decreasing gravity strength. Dividing the maximum thermal exposure at the exit by the hatch area gives the average heat flux through the exit. It makes sense to look at the heat flux since the opening area is not the same. 

Under normal gravity conditions, the maximum heat flux at the exit occurs for model 3 (802 kW/$\mathrm{m^2}$) followed by model 4 (270 kW/$\mathrm{m^2}$), 2 (264 kW/$\mathrm{m^2}$), and 1 (174 kW/$\mathrm{m^2}$) just as in the case of maximum impact force and maximum pressure. As expected, the contribution of convective heat transfer (the second term on the right-hand side of Eq.~(\ref{heat_flux})) plays an important role in increasing the thermal exposure consistently with the trends of the maximum pressure inside the enclosure and those of the impact force.

Fu et al.~\cite{fu_numerical_2014} found in their work that exposure to a constant heat flux of 7 kW/$\mathrm{m^2}$ for 20 s will cause second-degree skin burn in blackened living skin. As the heat flux increases, the time of exposure to cause the skin burn decreases. Second-degree skin burns occur almost instantaneously at heat flux values more than 50 kW/$\mathrm{m^2}$. It can be seen from Figure~\ref{heat_flow} that all heat flux values are well above 50 kW/$\mathrm{m^2}$. Thus, the thermal exposure due to backdraft is severe and is more important than the impact force.

\section{Conclusions}\label{conclusions}
A numerical investigation of the effect of gravity on backdraft phenomena in an enclosure has been conducted for four different configurations of opening geometry. It was investigated whether any visual cues could be obtained to predict the arrival of the deflagration wave through the hatch. The effects of backdraft in the form of impact force and thermal exposure through the exit were also assessed. The setup from the experiment was recreated with initial conditions set as noted in the experiment before the opening of the hatch. Soot was added in the methane reaction and in the enclosure to model soot formation during the incomplete combustion of methane. The conclusions drawn for the questions outlined in this work are as follows.

\begin{enumerate}
    \item Backdraft occurs under all gravity strengths despite variation in the time at which it is formed. This observation is consistent with the observations made by Ashok and Echekki~\cite{ashok_numerical_2021}. The results establish this observation for all four opening geometries.
    \item Ignition time decreases with the gravity strength in a nonlinear fashion. The trend of this time is consistent with the scaling of the gravity wave speed with gravity, as established by von Karman~\cite{von_karman_engineer_1940} and Benjamin~\cite{benjamin_gravity_1968}.
    \item We found that the time of onset of backdraft also exhibits a similar relationship, which indicates inverse scaling with the square root of the gravity constant.
    \item It is possible to use the amount of soot/smoke exiting the opening as precursor for the onset of the backdraft. For all the opening geometries studied in this work, we observe that there is just enough time to take precautions. This time decreases with the gravity strength.
    \item The damage caused by thermal exposure is more severe than that caused by the impact force at the exit.
    \item Some of the trends in the present results may be closely related to the maximum pressure established within the enclosure. These include the trends of the thermal exposure and the impact force. The opening geometry plays an important role in both the placement of the opening and its total area.
\end{enumerate}
As observed above, both opening placement and size have a strong impact on the timing of the backdraft and its consequences in terms of thermal exposure and the presence of precursors. These effects are present at different gravity conditions. The quantitative results are expected to be dependent on the enclosure size as well as the placement of the ignition source. Yet, they do provide an understanding of the trends and the mechanisms that drive them.
\printcredits

\section*{Declaration of competing interest}
The authors declare that they have no known competing financial interests or personal relationships that could have appeared to influence the work reported in this paper.

\section*{Acknowledgements}
The first author acknowledges the support of Mr. Brian Klein and the sales team from Thunderhead Engineering Inc. for providing multiple licenses of PyroSim for carrying out this research. The first author also acknowledges the computing resources provided by the North Carolina State University High-Performance Computing Services Core Facility (RRID:SCR\textunderscore02216) and the support from Dr. Jianwei Dian to set up the code in the HPC and error troubleshooting. The first author extends his gratitude to the developers of FDS Dr. Jason Flyod, Dr. Kevin McGrattan, Dr. Randy McDermott, Dr. Glenn Forney, and Dr. Jonathan Hodges for their guidance. The first author also thanks Mr. Shreyas G. Ashok for his support. This research did not receive any specific grant from funding agencies in the public, commercial, or not-for-profit sectors. 

\section*{Data availability statement}
The research data generated from this work can be made available on request from the first author.

\bibliographystyle{unsrtnat}

\bibliography{cas-refs}


\bio{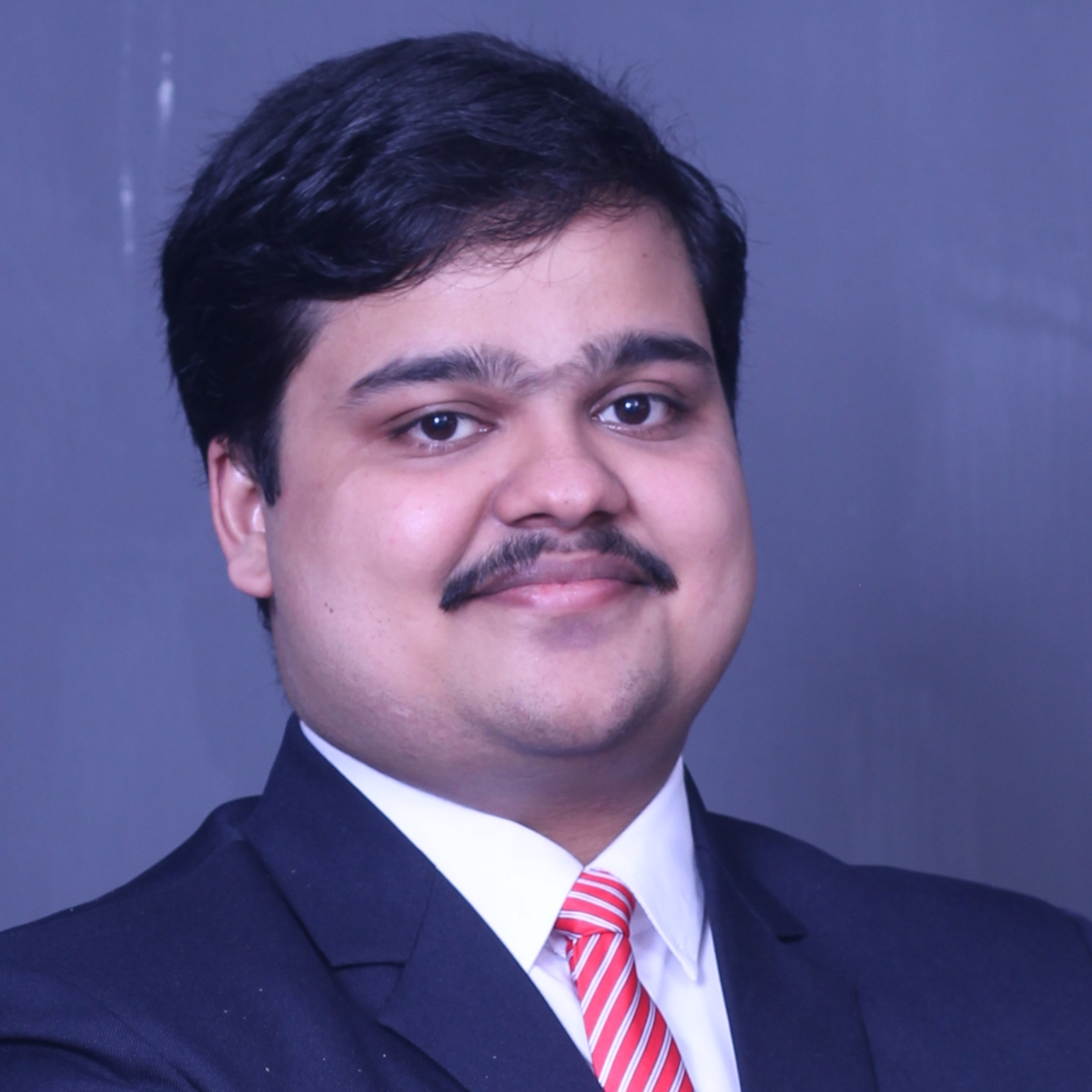}
Vijayananda Vivek Devananda received his BTech from Mahatma Gandhi University, Kerala, India in 2017. He completed his MS in aerospace engineering from NC State University under the supervision of Dr. Tarek Echekki in 2024. He is interested in researching combustion, propulsion, and optical diagnostics.
\endbio

\bio{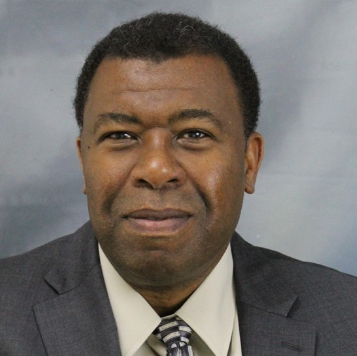}
Dr. Tarek Echekki has been a professor in the Department of Mechanical and Aerospace Engineering at NC State University since 2002. He received his Ph.D. in mechanical engineering from Stanford University in 1993. His interests are in combustion theory, DNS, and turbulent combustion modeling. Professor Echekki is a Fellow of the Combustion Institute and the American Society of Mechanical Engineers and an Associate Fellow of the American Institute of Aeronautics and Astronautics. He also serves as the Associate Editor for the ASME Journal of Heat Transfer.
\endbio

\end{document}